\documentclass[conference]{IEEEtran}

\usepackage{cite}
\usepackage[utf8]{inputenc}
\usepackage{amsmath}
\usepackage{graphicx}
\usepackage{color}
\usepackage{verbatim}
\usepackage{multirow}
\usepackage{array}
\usepackage{tabularx}
\usepackage{soul}
\usepackage{listings}
\usepackage{xcolor}
\usepackage{filecontents}
\usepackage{textcomp}
\usepackage{enumitem}
\usepackage{subcaption}

\input{prolog-language}
\usepackage[super]{nth}
\usepackage{hhline}
\usepackage{ragged2e}
\usepackage{url}

\newcommand{\etal}{\emph{et al. }}
\newcolumntype{M}[1]{>{\centering\arraybackslash}m{#1}}

\begin{document}

\title{Extending Attack Graphs to Represent Cyber-Attacks in Communication Protocols and Modern IT Networks}

\author{\IEEEauthorblockN{Orly Stan\IEEEauthorrefmark{1}, Ron Bitton\IEEEauthorrefmark{1}, Michal Ezrets\IEEEauthorrefmark{1}, Moran Dadon\IEEEauthorrefmark{1}, Yuval Elovici\IEEEauthorrefmark{1}, Asaf Shabtai\IEEEauthorrefmark{1}\\ Masaki Inokuchi\IEEEauthorrefmark{2}, Yoshinobu Ohta\IEEEauthorrefmark{2}, Yoshiyuki Yamada\IEEEauthorrefmark{2}, Tomohiko Yagyu\IEEEauthorrefmark{2}\\}
\IEEEauthorblockA{\IEEEauthorrefmark{1}Department of Software and Information Systems Engineering, Ben-Gurion University of the Negev}
\IEEEauthorblockA{\IEEEauthorrefmark{2}Security Research Laboratories, NEC Corporation}
}

\markboth{Journal of \LaTeX\ Class Files,~Vol.~14, No.~8, August~2019}%
{Shell \MakeLowercase{\textit{et al.}}: Bare Demo of IEEEtran.cls for IEEE Journals}

\maketitle

\begin{abstract}
An attack graph is a method used to enumerate the possible paths that an attacker can execute in the organization network.
MulVAL is a known open-source framework used to automatically generate attack graphs.
MulVAL's default modeling has two main shortcomings.
First, it lacks the representation of network protocol vulnerabilities, and thus it cannot be used to model common network attacks such as ARP poisoning, DNS spoofing, and SYN flooding.
Second, it does not support advanced types of communication such as wireless and bus communication, and thus it cannot be used to model cyber-attacks on networks that include IoT devices or industrial components.
In this paper, we present an extended network security model for MulVAL that: (1) considers the physical network topology, (2) supports short-range communication protocols (e.g., Bluetooth), (3) models vulnerabilities in the design of network protocols, and (4) models specific industrial communication architectures.
Using the proposed extensions, we were able to model multiple attack techniques including: spoofing, man-in-the-middle, and denial of service, as well as attacks on advanced types of communication.
We demonstrate the proposed model on a testbed implementing a simplified network architecture comprised of both IT and industrial components.
\end{abstract}


\section{Introduction}
Risk assessment is a process that enables system stakeholders to assess the risks to their system and select suitable countermeasures.
An attack graph is a risk assessment method used to enumerate the possible attack paths that an attacker can take in order to compromise an organization \cite{phillips1998graph}.
This is done by modeling the prerequisites and consequences of exploiting software vulnerabilities, as well as the attacker's potential lateral movements.
Using attack graphs, system stakeholders can identify potential attack paths, assess their risk, visualize an exhaustive attack surface, and plan efficient countermeasures.

Various types of attack graphs have been proposed in the literature, including state attack graphs \cite{sheyner2002automated}, dependency attack graphs \cite{Jajodia2005topological}, multi-prerequisite graphs \cite{ingols2006practical}, and logical attack graphs \cite{ou2006scalable}.
In this research, we focus on the logical attack graph: a directed and-or graph in which the \textit{root} represents an attacker's goal, \textit{nodes} represent subgoals (attack steps), and \textit{leafs} represent facts about the system. 
Both facts and subgoals can be disjunctively or conjunctively connected to the next subgoals or a final goal.

MulVAL, a logic-based network security analyzer~\cite{ou2005mulval}, is a well-known open-source framework for automatically constructing attack graphs.
MulVAL models the interactions of software vulnerabilities with system and network configurations, while automatically extracting information from formal vulnerability databases (such as the National Vulnerability Database -- NVD~\cite{nvd}) and network scanning tools (such as Nessus~\cite{nessus}).

While the MulVAL framework is a very powerful tool for generating attack graphs, it currently suffers from the following major limitations. 
First, it is broadly focused on application (software) vulnerabilities and does not properly incorporate weaknesses in network architectures and protocols.
For instance, ARP poisoning, DNS spoofing, and SYN flooding are network attacks that cannot be properly modeled using only application vulnerabilities.
Thus, MulVAL is likely to create an incomplete view of the possible attacks and attack paths in an organizational network.
Second, MulVAL is specifically designed to model traditional IT environments and does not support advanced communication types such as wireless and bus communication.
Thus, it cannot be used to model cyber-attacks on modern enterprise networks, such as those that include IoT or industrial components.

Previous attempts to augment MulVAL for representing vulnerabilities in network protocols did not provide a comprehensive (general) network model but addressed and modeled specific network attacks (e.g., ARP poisoning in \cite{acosta2016augmenting}).
In this paper we extend the MulVAL framework with a comprehensive network modeling.
The proposed extension considers the seven layers of the OSI network model (including the physical network topology), supports short-range communication protocols which are very common in IoT environments, and models specific industrial communication architectures.
Consequently, the extended version of MulVAL supports the modeling of various network attacks, including attacks on the physical layer (e.g., bus spoofing, WEP cracking, Bluetooth PIN cracking), data link layer (e.g., ARP poisoning), network layer (e.g., IP spoofing, ICMP flooding), and application layer (e.g., DNS spoofing).
We demonstrated the proposed modeling in a testbed implementing a simplified network architecture, which is comprised of both IT and industrial components.
Finally, in order to maintain an end-to-end attack graph generation process (that is provided by the MulVAL framework), we present our dedicated agent that was implemented for automatically scanning the network and translating the scan results into our extended modeling.

To summarize, the contributions of this paper are as follows.
First, we present \emph{comprehensive network modeling} that 
considers the seven layers of the OSI network model (including the physical network topology, wireless networks and bus architecture), which are currently not supported by the MulVAL framework.
Second, we \emph{model multiple attack scenarios} that are not addressed by current attack graph frameworks (e.g., spoofing and wireless attacks).
Third, we present the \emph{implementation of a dedicated agent} which is used to automatically collect network configurations and generate the input for the extended model.

\section{Preliminaries}
\subsection{The Evolution of Attack Graph Generation Methods}
\textbf{Forward exploration for attack graph generation.}
The use of attack graphs for security assessment started almost 30 years ago.
Fundamental studies in this domain include the works of Philips and Swiler~\cite{phillips1998graph,swiler2001computer}, which introduced the concept of graph-based network vulnerability analysis, and presented an automated tool for generating attack graphs.
Their tool constructs the attack graph using the \textit{forward exploration} method, starting from the initial state (the initial position of the attacker) to any goal state (target).
This approach, however, is not efficient in terms of complexity, since it evaluates all existing vulnerabilities, including those that are not relevant to the goal state; therefore, it is impractical for applying on large-scale networks.
\\
\\
\textbf{Model checking for efficient security assessment.}
Ritchey and Ammann~\cite{ritchey2000using} suggested a simple network security model (consisting of hosts and their vulnerabilities, network connectivity, current state of the attacker, and available exploits), on which they apply a model checking technique (model checker SMV), in order to evaluate the security of a system by considering the relationships between different hosts in a network.
Ritchey \etal~\cite{ritchey2002representing} extended the simple network security model to represent TCP/IP connectivity. 
Within their model, which is called topological vulnerability analysis, the representation of a host was extended to include network services and configurations, and the network connectivity model was extended to support link, transport, and application layer security.
Based on Ritchey's connectivity model, Jajodia \etal developed a framework which automates the vulnerability analysis process by deploying an open-source version of the Nessus scanner~\cite{nessus} and combining its reports with a global exploits knowledge base.

Model checking methods are more efficient than generating an attack graph using forward exploration.
However, their main drawback is that they can only represent one attack path from source to target, while attack graphs enumerate all possible attacks.
\\
\\
\textbf{Model checking for attack graph generation.}
To address this limitation, Sheyner \etal~\cite{sheyner2002automated} introduced the attack graph toolkit, which utilizes model checking tools for automatically generating attack graphs, thus improving the efficiency and completeness of the model compared to previous works.
Jha \etal also suggested to use model checking for attack graph generation~\cite{jha2002two}.
Similar to~\cite{sheyner2002automated}, the authors modified the model checker NuSMV in order to produce a complete attack graph.

However, for the construction of attack graphs, model checking algorithms doesn't scale well; for example, in the work of Sheyner \etal \cite{sheyner2002automated} the construction of attack graph for a network with five hosts and eight exploits took two hours and resulted in 5948 nodes and 68364 edges.
Ammann \etal suggested to use a layered model to improve the scalability~\cite{ammann2002scalable}. 
They modeled facts (privileges, connectivity, and vulnerabilities) as generic attributes (e.g., the attacker has ftp access to a specific host) and exploits as transformations that map a set of preconditions to a set of postconditions.
Then, they arranged the attributes in layers according to amount of exploits required to obtain them.

Nevertheless, the main shortcoming of the above studies is that none of them automatically integrates vulnerability and network connectivity information into the attack graph generation process, and therefore none of them can support the analysis of large-scale networks.
\\
\\
\textbf{End-to-end attack graph generation frameworks.} The first framework to provide automatic end-to-end attack graph generation and analysis is MulVAL~\cite{ou2005mulval}, a logic-based network security analyzer.
MulVAL leverages exiting vulnerability databases (e.g., the National Vulnerability Database -- NVD) and network vulnerability assessment tools (e.g., Nessus) to automatically derive the inputs for the attack graph generation process.
This is done by using Datalog as the modeling language, which makes the integration of vulnerability databases and security assessment tools straightforward.
Another benefit of Datalog over model checking tools is its efficiency.
The complexity for model checking tools to enumerate all possible attack paths is exponential in the size of the network \cite{ou2005mulval}. 
In contrast, Datalog programs are polynomial in the size of network.

Following MulVAL, Ingols \etal presented NetSPA, an end-to-end framework for attack graph generation~\cite{ingols2006practical}.
NetSPA improves the input generation process by automatically extracting information from additional data sources such as Check Point firewalls and the CVE repository~\cite{cve}.

In 2013, Yi \etal~\cite{yi2013overview} compared several academic attack graph generation tools (TVA, Attack Graph Toolkit, NetSPA, and MulVAL), as well as commercial software (Cauldron \cite{Jajodia2011cauldron,cauldron} which is based on TVA, Firemon~\cite{firemon} which is based on NetSPA, and Skybox~\cite{skybox}) and concluded that:
(1) MulVAL is the most extendable and scalable framework;
(2) TVA and NetSPA are also scalable, however they are not open-source;
and (3) commercial tools are more scalable and provide an intuitive graphical user interface, however they are not suitable for research.
Therefore, in this study, we focus on the MulVAL attack graph framework.
\color{black}
\subsection{\label{subsec:mulval-back}The MulVAL Framework}
\begin{figure}[t]
    \centering
    \includegraphics[scale=0.3]{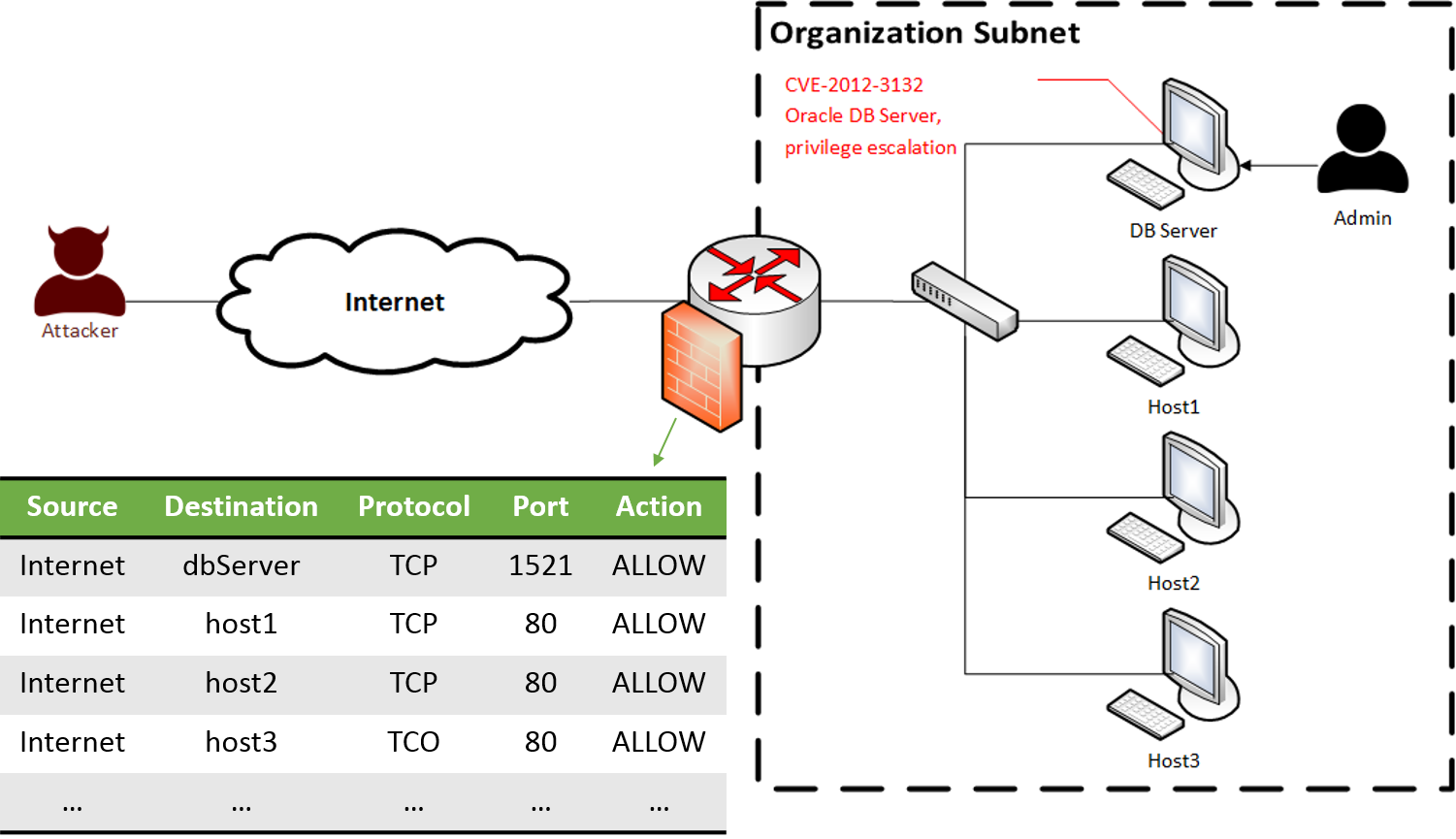}
    \caption{MulVAL example: code execution scenario.}
    \label{fig:mulval-example}
\end{figure}

MulVAL is a logic-based security analyzer that models the interactions between software bugs and network configurations (i.e., connectivity), while automating the information gathering process by deploying host-based scanners.
The modeling language used by MulVAL is Datalog~\cite{ou2005mulval}, which is a subset of the Prolog logic programming language.
We demonstrate MulVAL's modeling using the simple attack scenario illustrated in Figure \ref{fig:mulval-example}: the attacker aims to execute malicious code in the database server by abusing the free access from the Internet to the server and the vulnerable database software that allows privilege escalation (CVE-2012-3132).
\\
\\
\textbf{Datalog primitives and syntax.} The Datalog language consists of \emph{facts} and \emph{rules}, which are defined using \emph{predicates}. 
Predicate is an atomic formula of the form:
\[
    p(t_1,...,t_k)
\]
where, each $t_i$ can be either a constant (starting with a lower-case) or a variable (starting with an upper-case); wild cards are also supported and are marked with underscore ("\_"). For example, the following predicate states that: some vulnerability is present in the \textit{oracleDB} program running on \textit{dbServer}.

\begin{equation*}
        vulExists(dbServer, VulID, oracleDB).
\end{equation*}

Rules (referred to as \textit{interaction rules} in MulVAL) are represented using Horn clauses as follows:
\[
    P_0 :- P_1,...,P_n
\]
which essentially tells that if the predicates $P_1,...,P_n$ are true, then predicate $P_0$ is also true. 
The left part of the clause ($P_0$) is called the \textit{head} and the right part ($P_1,...,P_n$) is called the \textit{body}.
Facts are clauses with no body. For example, the following rule tells that a host can be remotely accessed if it runs a login service using specific port and protocol and it is accessible via the same protocol and port.
\begin{equation*}
\begin{split}
can&AccessHost(H) :- \\
    &\qquad	logInService(H, Protocol, Port), \\
    &\qquad	netAccess(H, Protocol, Port)).
\end{split}
\end{equation*}

\textbf{The XSB system.}
In order to execute a Datalog program, MulVAL uses the XSB environment, which is an extended implementation of the Prolog programming language that supports \emph{tabled execution}. 
Tabled execution prevents the recalculation of previously calculated facts (i.e., each fact is calculated only once). 
Since the number of facts is polynomial in the size of the network, executing a Datalog program has a polynomial time complexity.

\begin{figure}[t!]
    \hspace{25pt}
	\begin{subfigure}[h]{0.2\textwidth}
	    \includegraphics[scale=0.23]{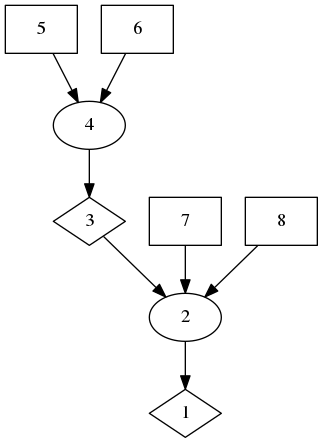}
	\end{subfigure}
	\hspace{-25pt}
	\begin{subfigure}[h]{0.2\textwidth}
		\tiny
		\begin{tabular}{|c|p{0.8\textwidth}|}\hline
		    Node & Description ($\mathcal{L}(Node)$)\\ \hline
    		1 & execCode(dbServer, root) \\ \hline
            2 & RULE 2: remote exploit of a server program \\ \hline
            3 & netAccess(dbServer, tcp, 1521) \\ \hline
            4 & RULE 6: direct network access \\ \hline
            5 & hacl(internet, dbServer, tcp, 1521) \\ \hline
            6 & attackerLocated(internet) \\ \hline
            7 & networkServiceInfo(dbServer, oracleDB, tcp, 1521, root) \\ \hline
            8 & vulExists(dbServer, 'CVE-2012-3132', oracleDB, remoteExploit, privEscalation) \\ \hline
		\end{tabular}
	\end{subfigure}
	\caption{MulVAL example: code execution attack graph.}
    \label{fig:mulval-example-ag}
\end{figure}

Attacks are simulated according to the \textit{attackerGoal(p)} facts (where $p$ is a predicate defined in the Datalog program) that are specified in the input to MulVAL (the complete example input is given in Appendix \ref{app:mulval-example-input}). 
The simulation is performed by querying the Datalog program for the predicate $p$; the attack graph is constructed based on the trace generated by this query. A detailed description of the Datalog trace analysis and attack graph construction is provided by Ou \etal in \cite{ou2006scalable}.
\\
\\
\textbf{Attack graph representation.}
MulVAL generates a logical attack graph as defined in \cite{ou2006scalable}: $AG = (N_r, N_p, N_d, E, \mathcal{L}, \mathcal{G})$, where $N_r, N_p, N_d$ are the sets of nodes (derivations, primitive facts, and derived facts correspondingly), $E$ is the set of edges, $\mathcal{L}$ is a mapping from a node to its label (i.e., the predicate it represents), and $\mathcal{G}$ is the node representing the attacker's goal. Figure \ref{fig:mulval-example-ag} presents the attack graph generated by the MulVAL tool\footnote{MulVAL tool: \url{http://www.arguslab.org/software/mulval.html}} for the attack scenario described above.

Derivation nodes (visualized as circles) correspond to interaction rules and represent the reason for a fact to become true, they can be viewed as attack steps ($N_r=\{2,4\}$).
Primitive facts (visualized as rectangles) correspond to the information given as input ($N_p=\{5,6,7,8\}$), while derived facts (visualized as diamonds) are the results of applying interaction rules on the primitive facts (can be viewed as attack step consequences; i.e., $N_d=\{1,3\}$).
An edge $e \in E$ can connect between a primitive or derived fact and a derivation ($e \in \{(N_p \cup N_d)\times N_r\}$), or between a derivation and a derived fact ($e \in \{N_r \times N_d\}$).
The derivation nodes imply an \textit{and} relation between their incoming nodes, which represent the preconditions for performing the corresponding actions; while the derived fact nodes imply an \textit{or} relation between their incoming nodes, which represent the various actions that result in the same consequence.

\section{Related Work}
The main limitation of MulVAL is that it only considers software vulnerabilities and does not address vulnerabilities in the design of network protocols.
Previous studies focused on extending the expressibility of MulVAL's security model. 
Appendix \ref{app:prev-work-comp} presents a comparison between MulVAL's modeling, previously suggested extensions, and our proposed extensions.

Froh \etal~\cite{bacic2006mulval, froh2009mulval} developed two extensions for MulVAL.
The first extension includes three main improvements: a more realistic network model that includes subnets and routers; modeling of data authorization policies; and an algorithm for prioritizing attack paths based on their impact on the organization.
The second extension includes modeling of additional safeguards (e.g., application level authentication) and a definition of high-level IT services for better attack graph analysis.
Although these extensions produce more realistic modeling, they are limited to software vulnerabilities and cannot be used to model vulnerabilities in network protocols or modern IT networks (such as those that include wireless communication). 

These works, however, focus on the efficiency and post-assessment of the attack graph and not on extending the framework to represent additional scenarios.

Acosta \etal~\cite{acosta2016augmenting} extended MulVAL to represent a specific data link vulnerability through which the ARP spoofing attack can be modeled.
In this study, we propose a comprehensive network model, which considers the seven layers of the OSI model, supports wireless and short-range communication protocols, and models specific industrial communication architectures.
Our proposed modeling enables the representation of various attacks on the physical, data link, network, and application layers.
Furthermore, we develop a prototype system that automatically collects network configurations, converts them to MulVAL's input, and generates the attack graph.

\section{\label{sec:network-model}Network Modeling}
\subsection{Overview}
In order to support more realistic network modeling, we added new predicates and interaction rules to MulVAL's original rule set (presented in~\cite{ou2005logic}).
These predicates can be categorized into: \textit{physical network topology, network communication, principal access, host configuration, and vulnerability}.

\begin{figure}
  \centering
  \includegraphics[scale=0.42]{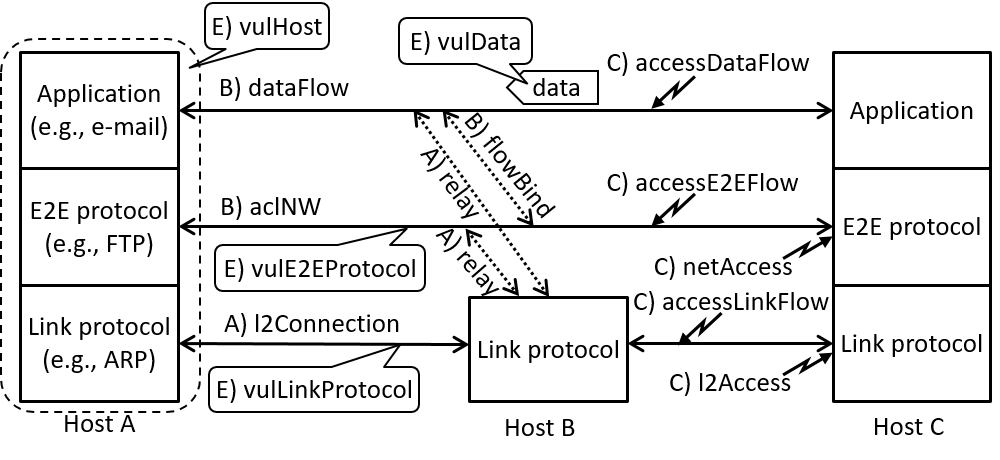}
  \caption{Illustration of the three-layer network structure and related predicates.}
  \label{fig:three-layer-model}
\end{figure}

In our extended modeling, IT networks are modeled using the following three layers: link, end-to-end, and application, as presented in Figure \ref{fig:three-layer-model}. 
The link layer integrates the network topology, users' actions, protocols, and vulnerabilities that are associated with a specific network zone (e.g., ARP spoofing in a subnet or WEP cracking). 
The end-to-end layer represents communication between specific hosts and services in remote hosts (OSI layers 3-7); it models the ability of two hosts to communicate using various protocols (e.g., FTP, HTTP) and the vulnerabilities existing in these protocols.
The application layer represents the user data and the data generated as a result of an end-to-end communication (e.g., the transmission of an e-mail).
To represent the connectivity between hosts in each layer, we defined the $l2Connection$, $aclNW$, and $dataFlow$ predicates. We also defined predicates that bind the different layers of communication.
The $flowBind$ predicate associates an end-to-end protocol with a data flow, and the $relay$ predicate associates a physical host with a data flow or an end-to-end communication. 
Since $l2Connection$ and $relay$ represent physical connections and components, they are categorized as \textit{physical network topology}; $aclNW$, $dataFlow$, and $flowBind$ are related to end-to-end and application communications, thus, are categorized as \textit{network communication}.
It is important to bind between the different layers because attacks on a lower layer can affect the upper layers of the communication.
For example, transmitting plain-text data over an encrypted end-to-end protocol can be considered safe; however, if an attacker succeed to break the end-to-end encryption he/she can read the transmitted data.

We also model principals' access to hosts in each layer.
The $l2Access$ and $netAccess$ predicates express a principal's access to a host in the link and end-to-end layers, respectively, while the $accessLinkFlow$, $accessE2EFlow$ and $acessDataFlow$ predicates express the principal's access to the corresponding layer's communication.

Vulnerabilities are classified into host, protocol, and data, according to their range of impact, and they are represented by the following predicates: $vulHost$, $vulLinkProtocol$, $vulE2EProtocol$, and $vulData$.  
A detailed description of the proposed extended network modeling follows.

\subsection{Physical Network Topology}
\textbf{Network Architecture and Components} (see Listing \ref{listing:network-components}). 
In order to bind an entity (e.g., a user or a host) to a network zone, MulVAL uses the \textit{located} predicate (line 1).
We find this coarse definition of zone insufficient for modeling the physical network topology.
Hence, we extend the \textit{located} predicate to specify the type ($Type$) of a network zone (line 2): an IP subnet to which $Host$ belongs ($ipSubnet$), a serial bus to which $Host$ is physically connected ($bus$), or the physical location (position) of $Host$.

\noindent To indicate the usage of a specific link layer protocol in a zone (e.g., serial Modbus in a bus) we introduce the \textit{existingProtocol} predicate (line 3). 
For example, $existingProtocol(subnet1,arp)$ indicates that hosts in $subnet1$ use ARP.

\noindent To be able to model network components (e.g., switch or router) in the physical network topology, we define the \textit{relay} predicates.
We distinguish between three types of network relays:
a link layer relay that controls the entire communication between $SrcHost$ and $DstHost$ (line 4);
a network layer relay that controls the communication of $SrcHost$ and $DstHost$ on a specific protocol ($Prot$) and port ($Port$) (line 5);
and an application layer relay that controls the communication of a specific application data flow (line 6).

We also define predicates that provide an indication about specific roles of network components (lines 7-11).
The $isGateway$ predicate indicates that $Router$ is the gateway of $Subnet$ (line 7).
The $isAP$ predicate provides an indication regarding an access point that broadcasts to a $WirelessRange$ and is connected to a $DstZone$ (line 8);
whereas, the $Prot$ argument specifies which protocol is used to establish the wireless connection (e.g., WEP), and the $SecurityConf$ argument specifies whether the access point uses a secured protocol ($SecurityConf=secured$) or is open to any connection request ($SecurityConf=open$). 
The $isNameResolver$ predicate indicates that $ResolverHost$ is the name resolver that is queried by $UserHost$ for finding the IP address associated with the name denoted by $Searched$ (line 9). \\
Finally, the $isMaster$ and $isSlave$ predicates are used to indicate master/slave components that are common in serial buses.
The master device issues commands to others (i.e., initiates the communication), and the slave components respond to the master's commands (lines 10-11).
It is important to distinguish between these two roles, since they affect the attacker's capabilities in a serial bus.

\lstinputlisting[
  style      = Prolog-pygsty,
  caption    = {Network components.},
  label = {listing:network-components},
  float=tp,
  floatplacement=tbp
]{network-component.pl}

\noindent \textbf{Network Connectivity} (see Listing \ref{listing:network-connectivity}). Within MulVAL, the communication between hosts is represented by the $hacl$ predicate, which represents an abstraction of some host's ability to reach another after considering the network's topology and security configurations \cite{ou2005logic} (e.g., if there is a firewall denying the communication between two hosts, an \textit{hacl} fact will not be generated).
We find this abstraction unsuitable for modeling the link and physical layers connections.
As mentioned, we distinguish between the link layer connectivity (which is not addressed by the original MulVAL modeling), and the end-to-end connectivity (which is originally addressed by the $hacl$ predicates). 
A link layer connectivity relates to the connections between hosts through a shared medium (e.g., Ethernet, Wi-Fi, Bluetooth, bus, etc.).
We represent this type of connectivity using the \textit{l2Connection} predicate (line 1), which indicates that two devices ($Dev1$ and $Dev2$) are connected via the same link ($LinkId$) of $Type$ and can communicate using protocol $Prot$. 
We consider the following three medium types:

\begin{enumerate}[leftmargin=12pt]

\item Ethernet: an Ethernet link ($ipSubnet$) enables the connectivity between two devices ($Dev1$ and $Dev2$) located in the same IP subnet (lines 2-5).

\item Bus: a serial link ($bus$) enables the connectivity between two devices ($Dev1$ and $Dev2$) connected to it (lines 6-9).

\item Wireless: a wireless link ($Type=wireless$) enables the connectivity between two devices that are located within the same wireless range ($WirelessRange$).
We support two wireless technologies: Wi-Fi and Bluetooth. \\
\textit{Wi-Fi} enables the connectivity between a wireless device ($Dev$) and an access point ($AP$).
We distinguish between an open access point (lines 10-12) and a secured access point (lines 13-16), in which the principal ($Principal$) is authenticated. 
When a host connects to an access point it becomes part of the subnet that the access point is connected to (denoted by $DstZone$); thus, we define the $located$ interaction rule (lines 17-20). \\
\textit{Bluetooth} enables the direct connection between two devices ($MasterDevice$ and $SlaveDevice$) using the Bluetooth technology ($Type=bluetooth$).
In this type of connection the two devices must be physically located within each other's range ($BluetoothRange$). 
In addition, the $SlaveDevice$ must be in discovery mode (lines 21-25).
\end{enumerate}

\lstinputlisting[
  style      = Prolog-pygsty,
  caption    = {Network connectivity.},
  label = {listing:network-connectivity},
  float=tp,
  floatplacement=tbp
]{network-connectivity.pl}

\subsection{Network Communication}

\textbf{Application Layer Communication} (see Listing \ref{listing:netCom}). 
In our proposed modeling, the application layer represents data exchange between two entities. 
We model the communication in this layer using the $dataFlow$ predicate (lines 1-2). 
This predicate associates an application layer communication ($FlowName$) with the two hosts ($SrcHost$ and $DstHost$) that generated it (line 1). 
For bidirectional communication, the $Direction$ parameter is set to $twoWay$; otherwise, it should be set to $oneWay$. 
Subsequently, in some cases it is sufficient to associate only one communication end with $FlowName$; thus, we also define a shorter version of the $dataFlow$ predicate (line 2), where $Host$ is either the source or destination of the communication represented by $FlowName$. 
Communication (data) flows are also associated with end-to-end protocols (e.g., the transmission of an email can be associated with SMTP).
To express this association, we define the $flowBind$ predicate (line 3). 

Some attack scenarios rely on the attacker's ability to exploit a specific type of data flow.
For example, in order to execute a Bluetooth PIN cracking attack, the attacker needs to capture the packets transmitted during the pairing process.
Thus, we include the definition of specific data flow types. An example is $isPairingProcess$ (line 4), which indicates that the data flow ($FlowName$) is the sequence of packets transmitted during a Bluetooth pairing process.

\lstinputlisting[
  style      = Prolog-pygsty,
  caption    = {Network communication.},
  label = {listing:netCom},
  float=tp,
  floatplacement=tbp
]{app-com.pl}

\noindent \textbf{Access Control} (see Listing \ref{listing:netCom}). MulVAL represents the network access control policy by $hacl$ predicates, which take the physical topology of the network into account \cite{ou2005logic}.
In our proposed modeling, we separate between the representation of physical topology and access control. 
We represent the network-based access control rules by the $aclNW$ predicate (line 6), which indicates that packets of protocol $Prot$, originating in host/subnet ($SrcHostOrsubnet$) and designated for a specific port ($Port$) in the host/subnet denoted by $DstHostOrsubnet$, are allowed. 
Host-based access control modeling will be described in Section \ref{subsec:host-conf-model}.

\subsection{Principal Access}
The human aspect must also be taken into account when modeling network access. 
To be consistent with our three-layer network modeling, we define a principal access predicate for each layer: host access, network access, and data access. 

\noindent \textbf{Host Access} (see Listing \ref{listing:principalHost}). Refers to the ability of a principal to login to a host.
This type of access is represented by the $localAccess$ predicate (line 1), which indicates that a principal ($Principal$) can access a host ($Host$) using the user account denoted by $User$. 
A principal access to a host can also be inferred from the following two scenarios: (1) an attacker who is located in a host has access to that host (lines 2-3), and (2) a principal that has local access to a host ($HostA$) can use it to gain access via the network to a remote host ($HostB$) by using a network login service and a previously obtained account (lines 4-9).

\lstinputlisting[
  style      = Prolog-pygsty,
  caption    = {Principal host access.},
  label = {listing:principalHost},
  float=tp,
  floatplacement=tbp
]{principalHost.pl}

\noindent \textbf{Network Access} (see Listing \ref{listing:principalNetwork}). Similar to the network topology and communication, we distinguish between the principal's direct access to a host through the link layer (e.g., access a component via a serial bus) and by using an end-to-end protocol (e.g., sending an email using SMTP).

Link layer access is represented by the $l2Access$ predicate (line 1) which indicates that a principal ($Principal$) can access $DstHost$ from $SrcHost$ via the medium denoted by $LinkID$. 
The $Type$ argument refers to the type of medium (similar to the $l2Connection$ predicate), and the $Prot$ argument specifies the protocol used in that medium (e.g., Modbus, WEP).
The ability of a principle to access a host via a shared communication medium is inferred by the interaction rule in lines 2-4: a principal ($Principal$) with local access to $SrcHost$ can access $DstHost$, assuming that they are both connected to a shared medium $LinkID$.

In the end-to-end layer, the ability of a principal to access hosts is represented by the $netAccess$ predicate (line 6), which indicates that $Principal$ can use $SrcHost$ to communicate with $DstHost$ on a specific port ($Port$) using the protocol $Prot$. 
This fact is inferred by the interaction rule in lines 7-10: a principal ($Principal$) with local access to $SrcHost$ can communicate with $DstHost$ on $Port$ using the protocol $Prot$ if this communication is enabled by the network access control mechanisms (represented by $aclNW$) as well as the $DstHost$'s access control mechanisms (represented by $aclH$).

\lstinputlisting[
  style      = Prolog-pygsty,
  caption    = {Principal network access.},
  label = {listing:principalNetwork},
  float=tp,
  floatplacement=tbp
]{principalNetwork.pl}

\lstinputlisting[
  style      = Prolog-pygsty,
  caption    = {Principal data access.},
  label = {listing:principalData},
  float=tp,
  floatplacement=tbp
]{principalData.pl}

\noindent \textbf{Data Access} (see Listing \ref{listing:principalData}). We distinguish between two types of data: data at rest and data in motion.
Access to data at rest is modeled in MulVAL by the $accessFile$ predicate (line 1), which indicates that $Principal$ can access files in $Path$ present in $Host$ with $AccessPerm$ permissions (read, write, or execute).
We extend MulVAL to model access to data in motion as well by defining the $accessDataFlow$ predicate (line 3), which indicates that $Principal$ has access to the data flow denoted by $FlowName$.
There are three possible access types:
(1) \textit{view} - the principal is able to view $FlowName$ but not necessarily read its content, (2) \textit{read} - the principal can read the content of $FlowName$, i.e., see it in plain text, and (3) \textit{write} - the principal can manipulate the data in $FlowName$.
These access types can contain each other: if a principal can read a data flow, he/she can also view it; and if a principal can write to a data flow, he/she can also read it.

The interaction rules in lines 4-8 and 9-14 describe the ability to view wireless traffic. 
Lines 4-8 describe a scenario in which a principal's host ($SideHost$) is located in the wireless range ($WirelessRange$) of the communication between $HostA$ and $HostB$. 
Based on the assumption that $SideHost$ is equipped with a network interface, $Principal$ can capture the wireless traffic in that range. 
Similarly, lines 9-14, describe a scenario in which the $SideHost$ is positioned in the broadcast range of a device (e.g., access point) that is a relay to the communication that the principal wishes to capture.

To be consistent with the three-layer network model, we also defined \textit{accessLinkFlow} and \textit{accessE2EFlow} predicates (lines 16-17) in order to represent a principal's access to the communication in these layers. We also included interaction rules to bind protocol vulnerabilities with the data access level they permit across the network layers.
These definitions are important, since actual attacks tend to be executed in this step-by-step manner, thus they make our model realistic and practical.

\begin{figure}[h]
  \centering
  \includegraphics[scale=0.45]{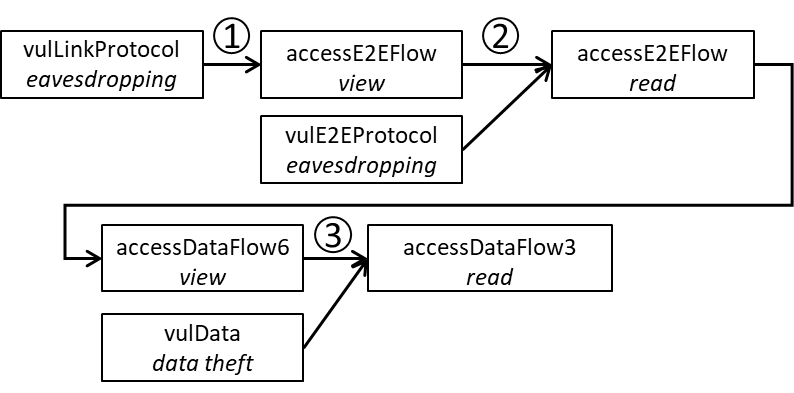}
  \caption{Inference structure of accessDataFlow with read access type}
  \label{fig:accessXXXFlow_inference}
\end{figure}

Figure \ref{fig:accessXXXFlow_inference} presents an example of this.
In the first step (denoted by \textcircled{1}), an attacker who can access the link layer (e.g., by connecting his/her laptop to a layer two switch) can try to exploit link layer vulnerabilities (specified by the $vulLinkProtocol$ predicate). If the attacker succeeds, he/she will be able to view the end-to-end layer flow (represented by $accessE2EFlow$ with the $view$ access type). 
In the next step (denoted by \textcircled{2}), the attacker can try to exploit an end-to-end protocol vulnerability (represented by the $vulE2EProtocol$ predicate). If the attacker succeeds, he/she can read the end-to-end layer flow (represented by $accessE2EFlow$ with the $read$ access type), which means that the attacker can view the data flow (represented by $accessDataFlow$ with the $view$ access type). 
In the final step (denoted by \textcircled{3}), the attacker can try to read the data flow (represented by $accessDataFlow$ with the $read$ access type) by attempting to exploit data vulnerabilities (represented by the $vulData$ predicate).

\subsection{\label{subsec:host-conf-model}Host Configuration}
To represent host-based attacks we need to be able to indicate the characteristics of programs (e.g., login service), data (e.g., credentials), and access control configuration that are present on the host. All of the predicates related to host configuration are presented in Listing \ref{listing:hostConf}.

\noindent \textbf{Data.} Some attack scenarios rely on the attacker's ability to access sensitive content. 
Therefore, we introduce specific predicates to represent special data types.
The $isCredential$ predicate indicates that the data denoted by $DataName$ is the credentials of $User$ on $Host$ (line 1).
This predicate can also be used to indicate that a data flow contains credentials (e.g., in the case of remote login).

\noindent Another sensitive data type relates to name resolvers (e.g., DNS servers), i.e., the binding between a naming and an IP address. 
This data is represented by the $isNameResolverRecord$ predicate (line 2), which is used in the modeling of the DNS spoofing attack (see subsection \ref{subsec:spoofing}).

\noindent In addition, we provide the $dataBind$ predicate (line 3) to associate data ($DataName$) with a specific path ($Path$) in a host ($Host$).

\noindent \textbf{Program.} MulVAL defines two types of programs: network services and login services.
A \textit{network service} is a program running under some user permissions on the host while listening on a specific port and communicates with other instances using a specific protocol (line 5).
A \textit{login service} is a program that provides a user the ability to access the host (line 6). \\
We introduce a third type of program, \textit{local service}, which represents any program running on a host, regardless of its ability to communicate over the network (line 7). This way we can model multistep attacks that exploit vulnerabilities in "offline" services (e.g., exploit a vulnerable PDF reader by providing it with a malformed file).

In addition, we include the $dependsOn$ (line 8) and $bugHype$ (line 9) predicates. The $dependsOn$ predicate indicates that a program (denoted by $Prog$) running on $Host$, depends on some library (denoted by $Library$). This predicate is important for representing vulnerabilities that come from third party libraries or programs.
 
The $bugHype$ predicate indicates that a software bug is present in the program (denoted by $Prog$) running on a host (denoted by $Host$). The meaning of the $Range$ and $Consequence$ arguments is similar to those specified in the vulnerability predicates, and are explained in the following Section.

\noindent \textbf{Access Control.} The host-based access control rules are represented by the $aclH$ predicate (line 11), which indicates that the local access control mechanism on $Host$ allows $User$ to communicate from the host/subnet denoted by $SrcHostOrsubnet$ with the host/subnet denoted by $DstHostOrsubnet$ on $Port$ using the protocol $Prot$.

\lstinputlisting[
  style      = Prolog-pygsty,
  caption    = {Host configuration.},
  label = {listing:hostConf},
  float=tp,
  floatplacement=tbp
]{hostData.pl}

\subsection{Vulnerability}
Vulnerabilities are modeled in MulVAL using the $vulExists$ and $vulProperty$ predicates \cite{ou2005logic}.
These predicates, however, do not provide a clear distinction between different types of vulnerabilities; thus, we define three vulnerability predicates according to the assets they affect: $host$, $network$, and $data$.

\noindent \textbf{Host} (see Listing \ref{listing:vulHost}). A host vulnerability (e.g., a vulnerability in a software running on the host) is represented by the $vulHost$ predicate (line 1) defining a specific vulnerability ($VulID$) in a program ($Prog$) that exists in $Host$. 
The $Range$ argument refers to the context in which the vulnerability can be exploited and can be set to $localExploit$, $adjacent$, or $remoteExploit$ - similar to the attack vector (AV) metric of CVSS.\footnote{https://www.first.org/cvss/}
We also include interaction rules that represent a host vulnerability due to an implementation bug (lines 2-3)
or dependency in a vulnerable library (lines 4-6), similar to their basic definition in MulVAL \cite{ou2005logic}.

\lstinputlisting[
  style      = Prolog-pygsty,
  caption    = {Host vulnerability.},
  label      = {listing:vulHost},
  float=tp,
  floatplacement=tbp
]{vulnerability-host.pl}

\noindent \textbf{Network} (see Listing \ref{listing:vulNetwork}). A network vulnerability (i.e., a vulnerability in a communication protocol) is represented by the \textit{vulLinkProtocol} (lines 1-2) and $vulE2EProtocol$ (lines 9-10) predicates. 
The $vulLinkProtocol$ predicate in line 1 defines a vulnerability in a link layer protocol denoted by $Protocol$ (e.g., ARP, Bluetooth, WEP) that is used in some communication medium ($LinkID$) and contains the vulnerability $VulID$ that can be exploited via $Range$, and results in $Consequence$ (e.g., eavesdropping, data falsification).
The predicate in line 2 has a similar meaning, except that it provides an indication of a directed communication between two specific hosts ($SrcHost$ and $DstHost$). \\
We also define the $vulLinkProtocol$ interaction rule in lines 3-7, which infers a vulnerable link layer communication between $SrcHost$ and $DstHost$ from the fact that they are connected to the same communication medium ($LinkID$), in which a vulnerable link layer protocol ($Protocol$) is used.
Note that this interaction rule only holds for $ipSubnet$ or $bus$ medium types. \\
The $vulE2EProtocol$ predicate represents the use of an end-to-end protocol denoted by $Protocol$ (e.g., HTTP, SMTP) in the communication between two hosts on port $Port$ which contains the vulnerability $VulID$ that can be exploited via $Range$ and results in $Consequence$.
The predicate in line 9 indicates that $Host$ is either the source or destination of the vulnerable communication, and the predicate in line 10 represents a directed vulnerable communication.

\lstinputlisting[
  style      = Prolog-pygsty,
  caption    = {Network vulnerability.},
  label      = {listing:vulNetwork},
  float=tp,
  floatplacement=tbp
]{vulnerability-net.pl}

\noindent \textbf{Data} (see Listing \ref{listing:vulData}). Representing vulnerable data (e.g., unencrypted/unsigned communication between two hosts or unencrypted files on a host) is important to avoid detection of false attack paths. 
For example, an encrypted sensitive document will be protected from an eavesdropping attacker even if it is being exchanged between two hosts over FTP.
We distinguish between vulnerable data at rest (e.g., a file stored on a host) and vulnerable data in motion (e.g., network traffic exchanged between two hosts). 

Vulnerable data at rest is represented by the $vulData$ predicate (line 1), where $Data$ (e.g., a file) has the vulnerability $VulID$ (e.g., unsigned) which can result in $Consequence$ (e.g., data falsification).
We also provide the ability to associate a vulnerability with a specific range in some host's memory by defining the $vulBind$ predicate (line 2). This predicate indicates that upon successful exploitation of $VulID$, all of the data stored in $Path$ in $Host$ will be affected (e.g., read by the attacker). 

Vulnerable data in motion (i.e., a data flow) is represented by the $vulFlow$ predicate (line 4), which indicates that $FlowName$ contains a vulnerability denoted by $VulID$ (e.g., unencrypted) that enables $Consequence$ (e.g., eavesdropping).
Since data flows are related to the communication between two hosts via the network, we infer their vulnerabilities from communication protocol vulnerabilities. 
We consider two vulnerability types: lack of encryption and lack of a signature. \\
The interaction rule in lines 5-10 describes a scenario of an insecure data flow between $SrcHost$ and $DstHost$ (represented by the $dataFlow$ and $flowBind$ predicates) that is exchanged using an insecure end-to-end protocol and an insecure link protocol (represented by the $vulE2EProtocol$ and $vulLinkProtocol$ predicates), while the payload itself (e.g., a file) is not encrypted (represented by the $vulData$ predicate). \\
Similarly, in lines 11-17 the flow is only vulnerable within a specific zone. 
This situation can occur when: (1) the flow is transmitted via a relay host (e.g., a router or proxy) that is connected to the same zone as $SrcHost$, in which a vulnerable link protocol that enables eavesdropping is used (represented by the $vulLinkProtocol$ predicate); (2) the flow is transmitted using a vulnerable end-to-end protocol that can be exploited from the relay host and enable eavesdropping (represented by the $vulE2EProtocol$ predicate). 
The symmetric case (where the vulnerability exists in the zone of $DstHost$) can be defined in the same manner. \\
The interaction rule in lines 18-22 describes the scenario of eavesdropping on an unencrypted link (e.g., bus). 
In order for data to be vulnerable to link eavesdropping, it should be unencrypted (represented by the $dataFlow$ and $vulData$ predicates) and it should also be transmitted via a vulnerable link protocol (represented by the $vulLinkProtocol$ and $flowBind$ predicates). \\
The interaction rule in lines 23-27 describes a scenario in which unsigned data (represented by the $dataFlow$ and $vulData$ predicates) is transmitted via a vulnerable end-to-end protocol that enables data falsification (represented by the $vulE2EProtocol$ and $flowBind$ predicates).

\lstinputlisting[
  style      = Prolog-pygsty,
  caption    = {Data vulnerability.},
  label      = {listing:vulData},
  float=tp,
  floatplacement=tbp
]{vulnerability-data.pl} 

\section{\label{sec:attack-model}Attack Modeling}
In this section, we demonstrate how the proposed network modeling can be used to model network attacks, including: spoofing attacks, denial of service attacks (DoS), man-in-the-middle (MITM) attacks, attacks on wireless communication, and attacks on bus network architecture.

\subsection{\label{subsec:spoofing}Spoofing Attacks}
We distinguish between two types of spoofing attacks: spoofing in the link layer, in which the attacker can spoof the entire communication between the two hosts, and spoofing in the end-to-end layer, in which the attacker can spoof a specific service.
These two scenarios are represented using two predicates: $spoofLinkHost$ (Listing \ref{listing:spoofArp}, line 1) and $spoofE2EHost$ (Listing \ref{listing:spoofDNS}, line 1), which specify that a principal ($Principal$) can impersonate $ImpersonatedHost$ and fool $FooledHost$ using the $SpoofingHost$ by utilizing a vulnerability in a link/end-to-end protocol.
The $Consequence$ argument provides an indication about the outcome of the spoofing attack:
(1) the traffic is routed via $SpoofingHost$, e.g., in an ARP spoofing scenario ($Consequence=trafficTheft$); 
(2) $Principal$ can only deceive $FooledHost$, e.g., in bus spoofing scenarios ($Consequence=deception$).
We modeled the ARP and DNS spoofing attacks using these two predicates.

\noindent \textbf{ARP Spoofing} (see Listing \ref{listing:spoofArp}).
ARP spoofing is a technique by which an attacker can route the traffic intended for some host to the attacker's machine.
The attacker achieves this by sending fake ARP messages to a local area network (LAN) in order to associate the attacker's MAC address with the IP address of the victim's host.
In order to perform an ARP spoofing attack, two preconditions must be satisfied: 
(1) the impersonated host must use the vulnerable ARP protocol;
and (2) the attacker must be able to access the impersonated host via the link layer.

In our modeling we consider two ARP spoofing scenarios:
a scenario in which the impersonated and fooled hosts reside in the same subnet (lines 3-5), and a scenario in which they reside in different subnets (lines 6-10).  
The two preconditions for performing this attack are represented in both scenarios by the $vulLinkProtocol$ and $l2Access$ predicates.

\lstinputlisting[
  style      = Prolog-pygsty,
  caption    = {ARP spoofing.},
  label = {listing:spoofArp},
  float=tp,
  floatplacement=tbp
]{spoof.pl}

\noindent \textbf{DNS Spoofing} (see Listing \ref{listing:spoofDNS}).
In our modeling, we consider three DNS spoofing scenarios: (1) a scenario in which the attacker spoofs as the name server and provides malicious binding for legitimate requests, (2) a scenario in which the attacker provides a fake response by utilizing a race condition (i.e., answers faster than the legitimate server), and (3) a scenario in which the attacker changes DNS records in the name server/host cache.

The first DNS spoofing scenario can be represented directly by the $spoofE2EHost$ interaction rules that result in $trafficTheft$, in which the argument $ImpersonatedHost$ matches the identifier of the DNS server.

The second DNS spoofing scenario is more complicated, since the attacker utilizes the response time of a higher DNS resolver in order to send a fake response to the querying DNS server, binding the attacker's host's IP address to the queried naming.
Hence, in order to execute this attack, the DNS server should communicate using a vulnerable version of the DNS protocol.
As can be seen in lines 3-6, this prerequisite is  
represented by the $isNameResolver$ and $netAccess$ predicates. 
Following the successful execution of this attack, the querying host ($FooledHost$) will initiate communication with the attacker's host (which can be $AttackerHost$ or any other host) instead of the real host.

The third DNS spoofing scenario can be executed by gaining access to the DNS server with admin privileges.
We modeled this scenario in lines 7-11; as can be seen, the attacker ($Principal$) can log in using local access ($localAccess$ predicates) to the DNS server and modify the DNS records to associate the attacker's host with a naming of his/hers choice.
The success of the attack depends on the ability of the fooled host to access the attacker's host through the network (expressed by the $netAccess$ predicate).
Alternatively, this scenario can be executed by gaining $write$ access permission to some path in the DNS server, which happens to contain a DNS record.
We modeled this scenario in lines 12-16.

\lstinputlisting[
  style      = Prolog-pygsty,
  caption    = {DNS spoofing.},
  label = {listing:spoofDNS},
  float=tp,
  floatplacement=tbp
]{spoof.pl}

\subsection{Denial of Service Attacks}
We extend MulVAL's \textit{dos} predicate in order to indicate who is performing the attack (see Listing \ref{listing:dos}, line 1) as well as to support host-based (lines 3-7) and network-based (lines 8-13) DoS attacks.
In a host-based DoS attack, the attacker $Principal$ (represented by the $malicious$ predicate) exploits his/her login access to $Host$ (represented by the $localAccess$ predicate) which runs a vulnerable service that allows DoS (represented by the $localService$ and $vulHost$ predicates).
In a network-based DoS attack, the attacker $Principal$ exploits his/her access to a network service running on $DstHost$ (represented by the \textit{netAccess}, \textit{aclH}, and \textit{networkService} predicates), which contains a vulnerability that enables DoS upon successful exploitation (represented by the $vulHost$ predicate).

\lstinputlisting[
  style      = Prolog-pygsty,
  caption    = {DoS attacks.},
  label = {listing:dos},
  float=tp,
  floatplacement=tbp
]{dos.pl}

\subsection{Man-in-the-Middle Attacks}
A man-in-the-middle (MITM) attack is a scenario in which the attacker manages to pose as a relay of the communication between two hosts. 
We distinguish between a MITM attack in the link layer (Listing \ref{listing:mitm}, lines 1-6), in which the entire communication between $SrcHost$ and $DstHost$ is routed through the attacker's host ($AttackerHost$); and a MITM attack in the end-to-end layer (Listing \ref{listing:mitm}, lines 8-12), in which only a specific application layer protocol (represented by the $Port$ and $Prot$ arguments) is routed through the attacker's host.

\lstinputlisting[
  style      = Prolog-pygsty,
  caption    = {MiTM attacks.},
  label = {listing:mitm},
  float=tp,
  floatplacement=tbp
]{mitm.pl}

\subsection{Attacks on Wireless Communication}
One of the main contributions of this work is the ability to model attacks on wireless communication.
Specifically, we model two common attacks on wireless protocols: WEP cracking and Bluetooth PIN cracking, both of which enable the attacker to derive the encryption key used for the secured communication.

\noindent \textbf{\label{subsec:wep-cracking}WEP Cracking} (see Listing \ref{listing:wepCrack}).
The Wired Equivalent Privacy (WEP) algorithm is based on the RC4 stream cipher which has weak key scheduling and uses a small initial vector (IV) of 24 bits \cite{fluhrer2001weakness}. 
In order to crack WEP's encryption, the following two preconditions must be satisfied: (1) the attacker must capture a large amount of legitimate packets transmitted to/from the access point; and (2) the access point must use the vulnerable WEP protocol.
Afterwards the attacker can crack the encryption key by performing offline analysis \cite{fluhrer2001weakness}.
These two preconditions are modeled by the $crackAPEncKey$ interaction rule (lines 1-6), which utilizes the $accessDataFlow$, $relay$, $isAP$, and $vulLinkProtocol$ predicates.

By cracking the encryption key, the attacker can (1) eavesdrop on all of the traffic broadcasted by the compromised access point; and (2) authenticate to the access point.
These two different consequences are represented by the $accessDataFlow$ and 
$isAuthenticated$ interaction rules (lines 8-22).

\lstinputlisting[
  style      = Prolog-pygsty,
  caption    = {WEP cracking attack.},
  label = {listing:wepCrack},
  float=tp,
  floatplacement=tbp
]{wepCrack.pl}

\noindent \textbf{\label{subsec:wpa2-key-reinstall}WPA2 Key Reinstallation.} (see Listing \ref{listing:wpa2-key-reinstall}).
The Wi-fi Protected Access II (WPA2) protocol is widely used for protecting wireless networks. The Key Reinstallation Attack on the WPA2 protocol, which exploits a vulnerability in the 4-way handshake and tricks the victim into reinstalling an already in-used key, was published in 2017 by Vanhoef and Piessens \cite{vanhoef2017key}.
The 4-way handshake is used for assuring that both client and access point has the correct credentials, and to negotiate an encryption key for encrypting future communication. 
The encryption key is installed in the client after receiving the \nth{3} handshake message. If no acknowledgment is received (i.e., the \nth{4} handshake message), the access point might retransmit the \nth{3} handshake message to handle messages loss. Upon receiving this message, the client reinstalls the same encryption key and resets the nonce, which is also used in the derivation of the encryption key. As a result, an already used encryption key with a previously used nonce is used to further encrypt packets. Then, the attacker can derive the WPA2's encryption protocols keystream by using an encrypted message with a known content (or other methods when no such messages are found). In short, the key reinstallation attack is performed by the following two steps: (1) The attacker becomes a MITM between the host and the access point (e.g., by channel-based MITM technique \cite{vanhoef2014advanced}); and (2) the attacker prevents the \nth{4} handshake message from reaching the access point, which causes the retransmission of the \nth{3} message.

This scenario is represented by the interaction rule in lines 1-9. Lines 1-4 and 9 describe that the attacker's host ($AttackerHost$) and the victim's host ($Host$) are located in the same zone, so that the attacker is able to perform the first step of the attack (MITM). The $flowBind$ predicate in line 5 suggests that a data flow is transmitted using a specific protocol; in addition, the $isCredential$ predicate in line 6 provides indication that this data flow is part of the protocol's handshake (which the attacker plans to interrupt). Lines 7-8 describe that the access point that defined the zone ($WirelessRange$), in which both the attacker's host and victim's host are locates, uses a vulnerable communication protocol version that allows the execution of the key reinstallation attack. Thus, enabling the attacker to decipher future communication between the victim's host and the access point.

Furthermore, the authors in \cite{vanhoef2017key} found an additional vulnerability in several WPA2 implementations that reinstall an all-zero encryption key upon retransmission of \nth{3} handshake message. The corresponding interaction rule is presented in lines 10-15. This rule uniquely described this version of the key reinstallation attack by requiring that the communication protocol will be vulnerable to key reinstallation attack and that the victim's host ($Host$) will run a vulnerable version of the process (wpa\_supplicant) that enables for reinstalling the all-zero key. The $accessLinkFlow$ in line 12 implies that the attacker ($Principal$) managed to crack the encryption of the communication protocol ($Protocol$) that is used to communicate with the access point; together with the requirement vulnerability requirement in line 14, the $accessLinkFlow$ predicate should represent the successful exploitation of the key reinstallation attack (i.e., the rule in lines 1-9 is derived). Note that since the attacker knows the actual key after performing this attack (sequence of zeros), this scenario is represented by a $CrackAPEncKey$ rather than an $acessLinkFlow$ predicate.

\lstinputlisting[
  style      = Prolog-pygsty,
  caption    = {WPA2 key reinstallation attack.},
  label = {listing:wpa2-key-reinstall},
  float=tp,
  floatplacement=tbp
]{wpa2.pl}

\noindent \textbf{\label{subsec:pin-cracking}Bluetooth PIN Cracking} (see Listing \ref{listing:pinCrack}).
In 2005, Shaked \etal \cite{shaked2005cracking} presented a method for cracking the shared secret (referred to as PIN code) of two devices establishing a Bluetooth communication (using offline analysis).
In order to execute the attack, the attacker must eavesdrop on the entire pairing process between the two devices.
This precondition is represented by the $crackPINCode$ interaction rule (lines 1-6).
As can be seen, the attacker $Principal$ (represented by the $malicious$ predicate) can read the messages exchanged by two devices (represented by the $accessDataFlow$, $dataFlow$, and $flowBind$ predicates) during their pairing process (represented by the $isPairingProcess$ predicate). 
After capturing the messages, the attacker can perform the offline analysis presented in \cite{shaked2005cracking}.
The consequence of this attack is that the attacker acquires the capability to decipher the communication between $SrcHost$ and $DstHost$; we represent this capability using the \textit{accessDataFlow} interaction rule (lines 8-13).

\lstinputlisting[
  style      = Prolog-pygsty,
  caption    = {PIN cracking attack.},
  label = {listing:pinCrack},
  float=tp,
  floatplacement=tbp
]{pinCrack.pl}

\subsection{\label{subsec:bus-attacks}Attacks on Bus Communication}

Another major contribution of our work is the definition of serial bus topology, which can be found in critical infrastructure. In this section, we describe how DoS and spoofing attacks can be applied to bus architecture and how they can be modeled using our suggested network modeling.

\noindent \textbf{Denial of Service.}
In bus topology, all of the devices are physically connected to the same physical medium in order to communicate. A DoS attack on a serial bus can be executed in several ways, e.g., a compromised device can continually broadcast data and cause collisions on the bus, or a compromised device can broadcast data when detecting a data transfer to/from a specific device in order to prevent it from requesting service or transmitting data (i.e., DoS for a specific device).
Therefore, just one of the devices connected to a bus needs to be compromised in order to cause a DoS for all of the other connected devices.
    
To represent these scenarios, we define the $dos$ interaction rule (Listing \ref{listing:busAttacks}, lines 1-3), which reflects the fact that the attacker $Principal$ (represented by the $malicious$ predicate) can utilize his/her local access (i.e., login or ability to execute code) to a device connected to a bus in order to deny service to any other device connected to that bus (represented by the $l2Access$ predicate).

\noindent \textbf{Master/Slave Spoofing.}
Spoofing attacks on serial protocols are largely feasible due to the lack of authentication (e.g., serial Modbus used in OT systems).
In our current modeling, we only consider master/slave buses. We distinguish between two spoofing scenarios: (1) a compromised device connected to the bus masquerades as the master and sends fake data and commands, and (2) a compromised device connected to the bus masquerades as a slave and sends false data.
We distinguish between these two cases, since the abilities of the attacker can be affected by the role of the compromised component (i.e., a master or slave). 
Since we represent serial bus communication, the bus spoofing attack scenarios are represented as $spoofLinkHost$ interaction rules.
    
The master spoofing scenario is represented by the interaction rule in Listing \ref{listing:busAttacks}, lines 5-9. In this scenario, the attacker $Principal$ can log in to the $AttackerHost$ (represented by the $localAccess$ predicate) which is connected to the same bus as $FooledHost$ (represented by the $l2Connection$ predicate). By exploiting a vulnerability in the communication protocol used by $ImpersonatesHost$ and $FooledHost$ (represented by the $vulLinkProtocol$ predicate), $Principal$ is able to impersonate as $ImpersonatesHost$, which is the master in the bus ($isMaster$).

The slave spoofing scenario is represented by the interaction rule in Listing \ref{listing:busAttacks}, lines 10-14, which is similar to the master spoofing scenario, except for the fact that $ImpersonatesHost$ is a slave in the bus (represented by the $isSlave$ predicate) and not the master.

\lstinputlisting[
  style      = Prolog-pygsty,
  caption    = {Attacks on serial bus.},
  label = {listing:busAttacks},
  float=tp,
  floatplacement=tbp
]{bus-attacks.pl}

\section{\label{sec:prototype}Prototype Implementation of Automatic Fact Generation Process}
The main advantage of MulVAL over other attack graph solutions is that it provides an end-to-end framework for attack graph generation.
In order to preserve this strength, we developed a prototype system which automatically collects relevant information from the target system and generates the facts about the system with respect to the new network modeling (presented in Section \ref{sec:network-model}).
This prototype implements the following four fact generation methods:

\noindent \textbf{Scan:} facts that are generated directly from scan results; i.e., there is a complete mapping between the predicate's arguments and data in the scan results. A summary of the various tools that are used by the prototype system is presented in Table \ref{tab:agent-info}.
It should be noted that the current version of the prototype implementation supports the scanning of IP networks only.

\noindent \textbf{Knowledge:} facts that are generated using a knowledge base. The knowledge base contains general and system specific information, such as program characteristics, system configuration, etc. 

\noindent \textbf{Assumption:} facts that are generated with default values (e.g., \textit{unknown} or a wildcard).

\noindent \textbf{Manual:} facts that are generated manually, based on information provided by the operator/security expert, such as \textit{attackerLocated, attackGoal} and \textit{malicious(attacker)}.

A complete mapping of the four methods to predicates is presented in Appendix \ref{app:fact-gen}.

\begin{table*}[th]
    \caption{Information collected by the agent}
    \label{tab:agent-info}
    \scriptsize \centering
    \renewcommand*{\arraystretch}{1.3}
        \begin{tabular}{|c|l|l|}
            \hline
            \textbf{Type} & \textbf{Information collected} & \textbf{Tools (Windows/Linux)} 
            \\ \hline
            Topology & IP address, netmask, etc. & nmap 
            \\ \hline
            OS information & Type and version & nmap 
            \\ \hline
            \multirow{2}{*}{\centering Software and services} & Network: port number, protocol, service name, etc. & nmap, netstat \\ \cline{2-3}
                & Local: ID, name, executable, user, etc. &  WMI / dpkg-query, rpm, ps (login)
            \\ \hline
            Users and groups & ID, name, type, access permissions, etc. & WMI/passwd or sudo file (login) 
            \\ \hline
            Firewall & Host-based firewall rules & netstat firewall, advfirewall/ iptables-save, iptables-xml 
            \\ \hline
            Vulnerability & CVE-ID, CVSS score, CWE, related CPEs, etc. & OpenVAS, NVD 
            \\ \hline
        \end{tabular}
\end{table*}

\section{\label{sec:eval}Evaluation}
\begin{figure}[t]
    \centering
    \includegraphics[scale=0.3]{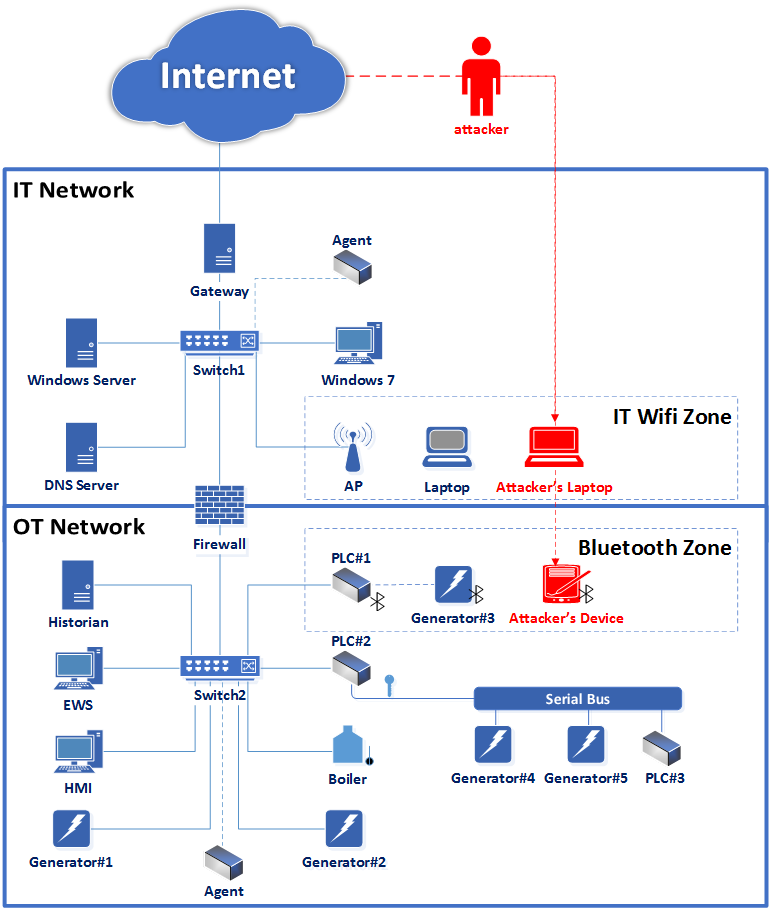}
     \caption{Testbed network topology}
    \label{fig:testbed-topology}
\end{figure}

In order to evaluate the application of our proposed modeling in a real environment, we established an operational testbed simulating a simple thermal power plant process (see Figure \ref{fig:testbed-topology}).

The testbed consists of: five generators, a boiler, a control panel, three programmable logic controllers (PLCs), an engineering work station (EWS), a historian, a human machine interface (HMI), a Windows workstation, a Windows server, a DNS server, an access point, and a laptop.
There are two subnets: \textit{IT Network} (the enterprise environment) and \textit{OT Network} (the OT environment) which are connected by a firewall. \textit{IT Network} is also connected to the Internet.
\textit{Bluetooth Zone} defines the range in which \textit{PLC\#1} can communicate with \textit{Generator\#3} using Bluetooth. \textit{IT Wifi Zone} is defined by the access point (denoted by \textit{AP}), which provides access to \textit{IT Network} for each authorized wireless device (e.g., the laptop) within \textit{IT Wifi Zone}. \textit{DNS Server} is the name resolver in \textit{IT Network}, which uses a vulnerable DNS protocol and accepts incoming communication from any host in \textit{IT Network}.

The enterprise hosts, HMI, EWS, and historian are virtual machines, that run a Windows Operating System, except for the historian which runs Linux. The behavior of the boiler and generators is simulated in the PLCs.
In order to collect real data from the testbed, we connect an instance of the prototype system to each network switch (denoted by \textit{agent} in Figure \ref{fig:testbed-topology}).

To ease readability, the graphs' nodes are marked as follows: blue nodes are associated with the system's topology and connectivity; red nodes are associated with vulnerabilities and the attacker's state; orange nodes are associated with software and services; and purple nodes are associated with data flows.

In this section, we evaluate the attack scenarios presented in Section \ref{sec:attack-model}.

\subsection{ARP Spoofing Attack}
\begin{figure}[h!]
	\begin{subfigure}[h]{0.2\textwidth}
	    \includegraphics[scale=0.20]{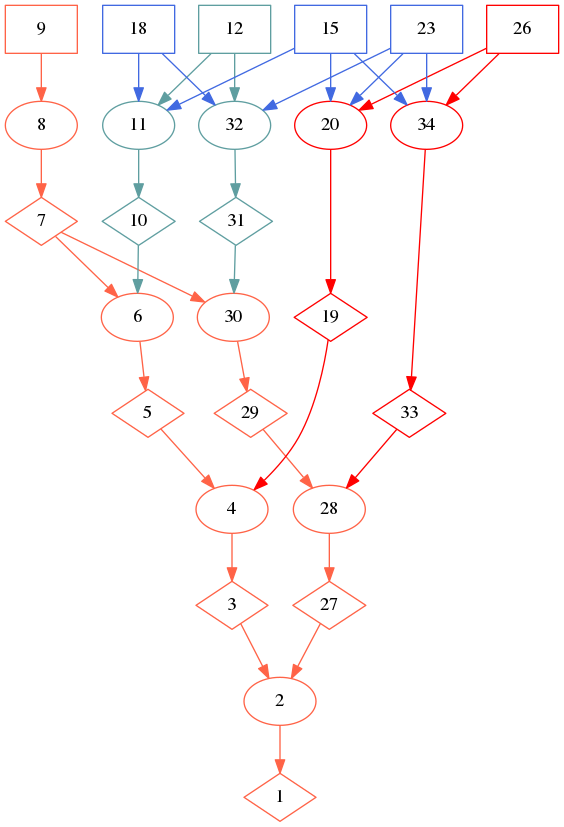}
	\end{subfigure}
    \hspace{5pt}
	\begin{subfigure}[h]{0.2\textwidth}
		\tiny
		\begin{tabular}{|c|p{0.9\textwidth}|}\hline
			1-2 & mitmLink(attacker, 'EWS','PLC1', 'HMI') \\ \hline
			3-4 & spoofLinkHost(attacker, 'PLC1', 'EWS', 'HMI', trafficTheft) \\ \hline
			5-6 & l2Access(attacker, 'HMI', 'PLC1', arp, 'OT Network',ipSubnet) \\ \hline
			7-8 & localAccess(attacker, 'HMI', admin) \\ \hline
			9 & attackerLocated('HMI') \\ \hline
			10-11 & l2Connection('HMI', 'PLC1', 'OT Network', arp,  ipSubnet) \\ \hline
			12 & existingProtocol('OT Network', arp) \\ \hline
			15 & located('PLC1', 'OT Network', ipSubnet) \\ \hline	
			18 & located('HMI', 'OT Network', ipSubnet) \\ \hline
			19-20 & vulLinkProtocol('EWS', 'PLC1', arpSpoofing, arp, adjacent, impersonateDst) \\ \hline
			23 & located('EWS', 'OT Network', ipSubnet) \\ \hline
			26 & vulLinkProtocol('OT Network', arpSpoofing, arp, adjacent, impersonateDst) \\ \hline	
			27-28 & spoofLinkHost(attacker, 'EWS', 'PLC1', 'HMI', trafficTheft) \\ \hline
			29-30 & l2Access(attacker, 'HMI', 'EWS', arp, 'OT Network', ipSubnet) \\ \hline
			31-32 & l2Connection('HMI', 'EWS', 'OT Network', arp, ipSubnet) \\ \hline
			33-34 & vulLinkProtocol('PLC1', 'EWS', arpSpoofing, arp, adjacent, impersonateDst) \\ \hline
		\end{tabular}
	\end{subfigure}
	\caption{MITM attack graph.}
    \label{fig:mitm-poc}
\end{figure}

Commonly used in IP networks, ARP is also used in the OT Network.
For simplicity and readability of the attack graph, we assume that the attacker has already managed to gain access to the HMI in order to execute the attack. 
After placing him/herself in the OT Network, the attacker can further exploit the use of ARP to impersonate any of the components in the subnet and eventually obtain a MITM position in the communication between the components.
The attack graph presented in Figure \ref{fig:mitm-poc} 
illustrates the attacker's steps in this attack scenario, where the attacker's goal is to become a MITM in the communication between \textit{EWS} and \textit{PLC\#1}.
Since the attacker is located in \textit{HMI} (nodes 7-9), which resides in the OT network as \textit{EWS} and \textit{PLC\#1} (nodes 15, 18, 23), he/she can access them both utilizing ARP that is used in this subnet (nodes 5-6, 10-12, 29-32).
Then, by exploiting the vulnerable ARP bidirectional communication between \textit{EWS} and \textit{PLC\#1} (nodes 19-20, 26, 33-34), the attacker is able to impersonate both \textit{EWS} and \textit{PLC\#1} (nodes 3-4, 27-28).
Finally, as a result of this bidirectional spoofing which causes the entire communication between \textit{EWS} and \textit{PLC\#1} to be routed via the attacker, we infer that the attacker is a MITM (nodes 1-2).

\subsection{DNS Spoofing Attack}
In this scenario, an attacker that placed his/her host in the \textit{IT Wifi Zone}, manages to connect to the \textit{IT Network} via \textit{AP}.
Then, the attacker exploits the vulnerable DNS protocol used by \textit{DNS Server} to poison its cache and bind \textit{Windows Server}'s naming to the attacker's IP address; the next time that someone queries for \textit{Windows Server}'s address, the attacker's IP will be returned, and the user will communicate with the attacker thinking it is really \textit{Windows Server}.

\begin{figure}[!h]
	\begin{subfigure}[h]{0.2\textwidth}
	    \includegraphics[scale=0.2]{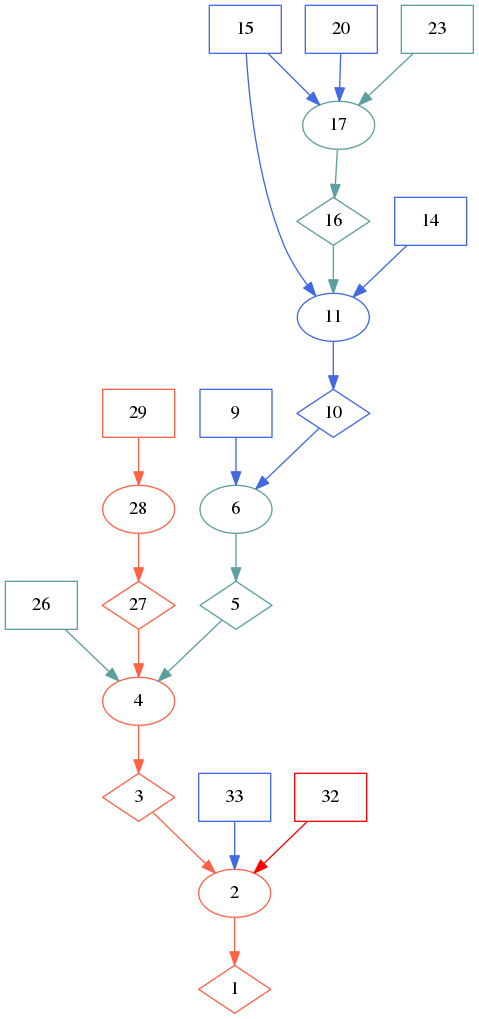}
	\end{subfigure}
	\begin{subfigure}[h]{0.2\textwidth}
		\tiny
		\begin{tabular}{|c|p{0.9\textwidth}|}\hline
			1-2 & spoofE2EHost(attacker, 'Windows Server', 'Windows 7', 'Attacker Laptop', \_, \_, trafficTheft) \\ \hline
			3-4 & netAccess(attacker, 'Attacker Laptop', 'DNS Server', dns, 53) \\ \hline
			5-6 & aclNW('Attacker Laptop', 'DNS Server', dns, 53) \\ \hline
			9 & located('DNS Server', 'IT Network', ipSubnet) \\ \hline
			10-11 & located('Attacker Laptop', 'IT Network', ipSubnet) \\ \hline
			14 & located('AP', 'IT Network', ipSubnet) \\ \hline
			15 & isAP('AP', 'IT Wifi Zone', 'IT Network', wep, secured) \\ \hline
			16-17 & l2Connection('Attacker Laptop', 'AP', 'IT Wifi Zone', wep, wireless) \\ \hline
			20 & located('Attacker Laptop', 'IT Wifi Zone', physical) \\ \hline
			23 & isAuthenticated(attacker, 'Attacker Laptop', 'AP') \\ \hline
			26 & aclH('Attacker Laptop', admin, 'Attacker Laptop', 'DNS Server', dns, 53) \\ \hline
			27-27 & localAccess(attacker, 'Attacker Laptop', admin) \\ \hline
			29 & attackerLocated('Attacker Laptop') \\ \hline
			32 & vulE2EProtocol('DNS Server', 'Windows 7', dnsCachePoisoning, dns, 53, remoteExploit, dataFalsification, twoWay) \\ \hline
			33 & isNameResolver('DNS Server', 'Windows 7', 'Windows Server') \\ \hline
		\end{tabular}
	\end{subfigure}
	\caption{DNS spoofing attack graph.}
    \label{fig:dns-spoof-poc}
\end{figure}

The attack graph generated for this scenario is presented in Figure \ref{fig:dns-spoof-poc}. 
The attacker can become a part of \textit{IT Network} (nodes 10-11) by  connecting to \textit{AP} (nodes 15-17, 20, 23), which is linked to that subnet. In so doing, the attacker can communicate with \textit{DNS Server} (nodes 3-4), since they are both connected to the same subnet (nodes 5-6, 9-10), and the attacker controls his/her device's configuration (nodes 26-29).
Then, the attacker can exploit the vulnerability in the DNS protocol (nodes 32-33) and execute the DNS spoofing attack.

\subsection{\label{subsec:syn-flood-eval}SYN Flooding Attack}
SYN flooding is a network-based DoS attack in which an attacker starts multiple TCP connections with the target host but doesn't complete them. In the vulnerable TCP implementation, the target host saves all of the half-open connections, which eventually exhausts the host's resources and makes it unresponsive to other connections.

\begin{figure}[!h]
	\begin{subfigure}[h]{0.2\textwidth}
	    \includegraphics[scale=0.23]{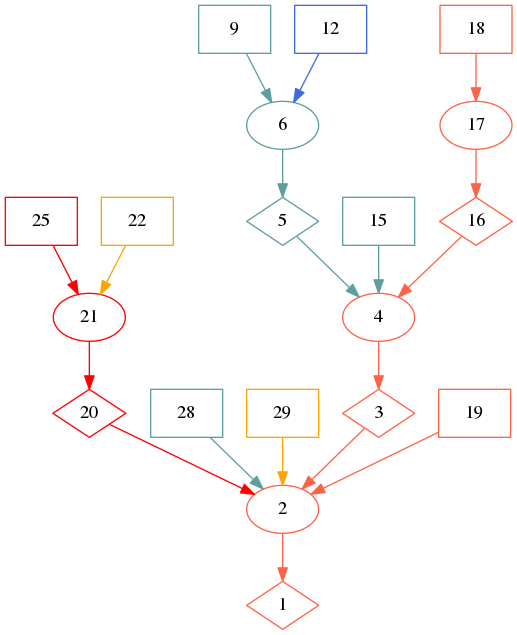}
	\end{subfigure}
    \hspace{15pt}
	\begin{subfigure}[h]{0.2\textwidth}
		\tiny
		\begin{tabular}{|c|p{0.9\textwidth}|}\hline
			1-2 & dos(attacker, 'Historian') \\ \hline
			3-4 & netAccess(attacker, 'Windows Server', 'Historian', tcp, 22) \\ \hline
			5-6 & aclNW('Windows Server', 'Historian', tcp, 22) \\ \hline
			9 & aclNW('Windows Server', 'OT Network', tcp, 22) \\ \hline
			12 & located('Historian', 'OT Network', ipSubnet) \\ \hline
			15 & aclH('Windows Server', admin, 'Windows Server', 'Historian', tcp, 22) \\ \hline
			16-17 & localAccess(attacker, 'Windows Server', admin) \\ \hline
			18 & attackerLocated('Windows Server') \\ \hline
			19 & malicious(attacker) \\ \hline
			20-21 & vulHost('Historian', synFlood, ssh, remoteExploit, dos) \\ \hline
			22 & dependsOn('Historian', ssh, tcp) \\ \hline
			25 & vulHost('Historian', synFlood, tcp, remoteExploit, dos) \\ \hline
			28 & aclH('Historian', \_, 'Windows Server', 'Historian', tcp, 22) \\ \hline
			29 & networkService('Historian', ssh, tcp, 22, \_) \\ \hline
		\end{tabular}
	\end{subfigure}
	\caption{SYN flooding attack graph.}
    \label{fig:syn-flood-poc}
\end{figure}

Within the testbed, the \textit{Historian} supports remote connection via SSH, which relies on TCP, and contains a vulnerable TCP implementation that allows the execution of the SYN flooding attack.
In this scenario we assume that the attacker managed to gain access to \textit{Windows Server} and execute the attack from it.
The attack graph for this scenario is presented in Figure \ref{fig:syn-flood-poc}. 
The attacker, who is located on \textit{Windows Server} (nodes 16-18), can utilize it to start a TCP connection  with \textit{Historian} on port 22 (nodes 3-4), since both the network-based and the host's host-based firewalls allow this connection (nodes 5-6, 9, 12 15).
The SSH service executed by \textit{Historian} (node 29) is vulnerable to the SYN flooding attack (nodes 20-22, 25), and together with the facts that (1) \textit{Historian} allows incoming communication from \textit{Windows Server} on port 22 (node 28), and (2) the attacker is able to initiate this communication (nodes 3-4), we conclude that the attacker can perform an SYN flooding attack on \textit{Historian} (nodes 1-2).

\subsection{WEP Cracking Attack}

In this scenario, the attacker physically places his/her host in the range of the access point (i.e., in \textit{IT Wifi Zone}), which is connected to \textit{IT Network}.
Based on the assumption that the attacker's host is adequately equipped to execute this attack, we infer that the attacker can capture the wireless signals transmitted in \textit{IT Wifi Zone}.
The access point uses WEP, which has a known vulnerability that allows key extraction, as its security protocol. 
One of \textit{Laptop}'s uses is to browse emails which results in communication with \textit{Windows Server}.

\begin{figure}[t]
	\begin{subfigure}[h]{0.5\textwidth}
	    \includegraphics[scale=0.2]{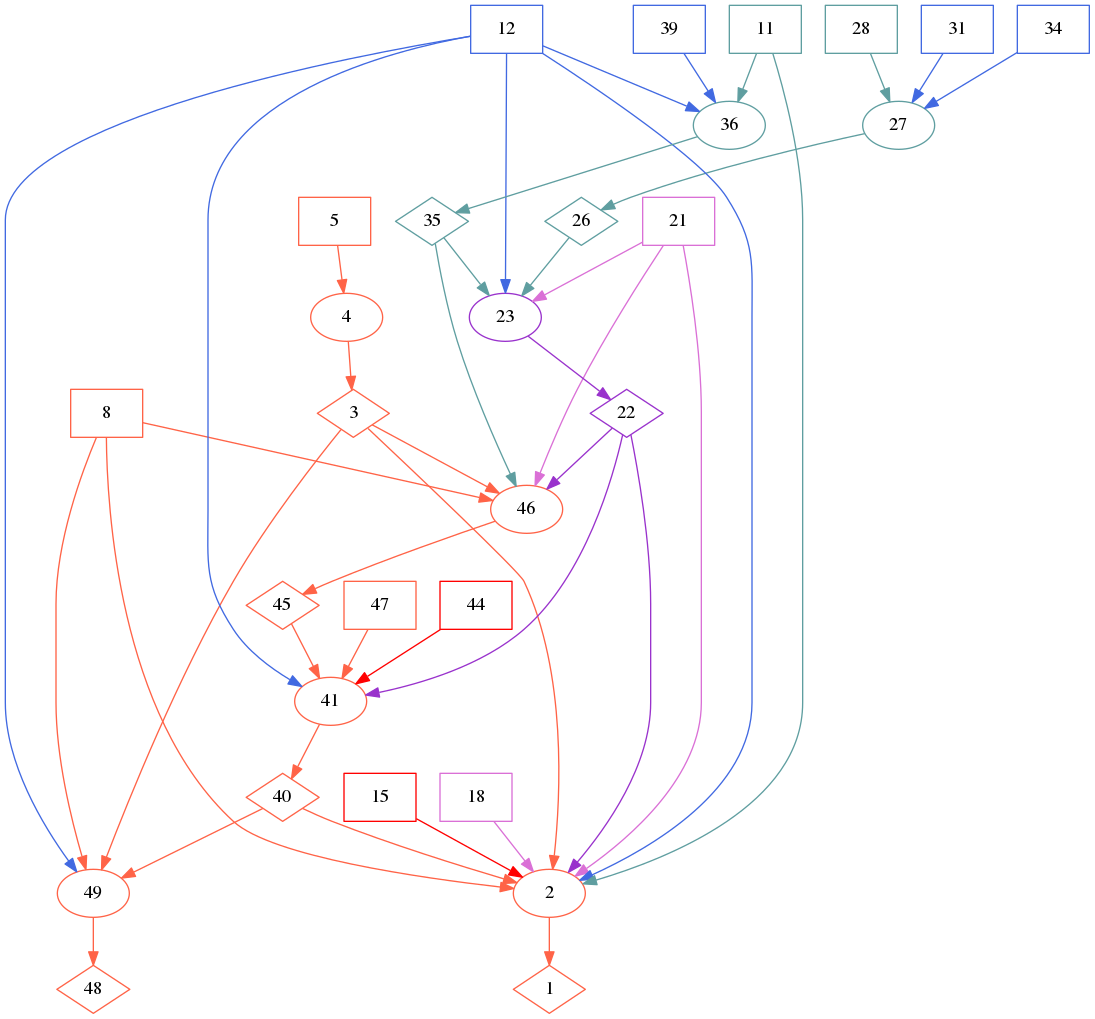}
	\end{subfigure}
	\begin{subfigure}[h]{0.5\textwidth}
		\tiny
		\begin{tabular}{|c|p{0.75\textwidth}|}\hline
			1-2 & accessDataFlow(attacker, emailFlow, read) \\ \hline
			3-4 & localAccess(attacker, 'Attacker Laptop', admin) \\ \hline
			5 & attackerLocated('Attacker Laptop') \\ \hline
			8 & located('Attacker Laptop', 'IT Wifi Zone', physical) \\ \hline
			11 & isAuthenticated('LaptopUser', 'Laptop', 'AP') \\ \hline
			12 & isAP('AP', 'IT Wifi Zone','IT Network', wep, secured) \\ \hline
			15 & vulE2EProtocol('Laptop','Windows Server', unencrypted, smtp, 25, remoteExploit, eavesdropping, twoWay) \\ \hline 
			18 & flowBind(emailFlow, smtp, 25) \\ \hline 
			21 & dataFlow('Laptop', 'Windows Server', emailFlow, twoWay) \\ \hline
			22-23 & relay('AP', emailFlow) \\ \hline 
            26-27 & l2Connection('AP', 'Windows Server', 'IT Network', arp, ipSubnet) \\ \hline
            28 & existingProtocol('IT Network', arp) \\ \hline
            31 & located('Windows Server', 'IT Network', ipSubnet) \\ \hline
            34 & located('AP', 'IT Network', ipSubnet) \\ \hline
            35-36 & l2Connection('Laptop', 'AP', 'IT Wifi Zone', wep, wireless)  \\ \hline
            39 & located('Laptop', 'IT Wifi Zone', physical) \\ \hline
            40-41 & crackAPEncKey(attacker, 'AP') \\ \hline
            44 & vulLinkProtocol('IT Wifi Zone', weakEncryption, wep, remoteExploit, keyExtraction) \\ \hline
            45-46 & accessDataFlow(attacker, emailFlow, view) \\ \hline
            47 & malicious(attacker) \\ \hline
            48-49 & isAuthenticated(attacker, 'Attacker Laptop', 'AP') \\ \hline
		\end{tabular}
	\end{subfigure}
	\caption{WEP cracking attack graph.}
    \label{fig:wep-crack-poc}
\end{figure}

The attack graph for this scenario is presented in Figure \ref{fig:wep-crack-poc}. 
As can be seen, \textit{AP} connects \textit{IT Network} with \textit{IT Wifi Zone} (node 12).
Since \textit{Windows Server} is connected to \textit{AP} in \textit{IT Network} (nodes 26-28, 31, 34), and \textit{Laptop} can connect to \textit{AP} in \textit{IT Wifi Zone} (nodes 11-12, 35-36, 39), we can infer that \textit{AP} is a relay to the communication between \textit{Laptop} and \textit{Windows Server} (nodes 21-23).
Since the attacker physically located his/her device in \textit{IT Wifi Zone} (nodes 3-5 and 8), the attacker can capture the communication between \textit{Laptop} and \textit{Windows Server} that is transmitted in the wireless zone (nodes 45-46).
Now, the attacker can exploit WEP's the weak encryption, which is used by \textit{AP} (node 44), and perform offline analysis on the captured traffic in order to extract \textit{AP}'s encryption key (nodes 40-41).
There are two consequences of this attack: (1) the attacker's device is authenticated by \textit{AP}, thus he/she is able to communicate freely with other hosts in \textit{IT Network} (nodes 48-49); (2) the attacker can decipher the communication between any host and \textit{AP} (nodes 1-2, 18).

\subsection{Bluetooth PIN Cracking Attack}
In this attack scenario, the attacker uses his/her device in order to capture the communication between two other Bluetooth components, which enables the attacker to perform the PIN cracking attack described in Subsection \ref{subsec:pin-cracking}. 

\begin{figure}[t]
	\begin{subfigure}[h]{0.2\textwidth}
	    \includegraphics[scale=0.2]{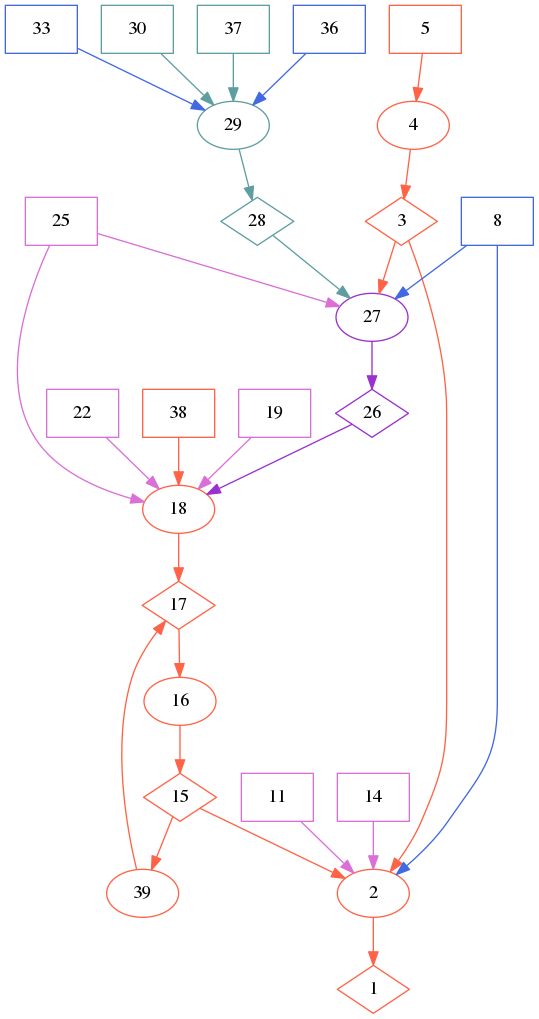}
	\end{subfigure}
	\hspace{2pt}
	\begin{subfigure}[h]{0.2\textwidth}
		\tiny
		\begin{tabular}{|c|p{1.02\textwidth}|}\hline
			1-2 & accessDataFlow(attacker, statusUpdate, read) \\ \hline
			3-4 & localAccess(attacker, 'Attacker Device', admin) \\ \hline
			5 & attackerLocated('Attacker Device') \\ \hline
			8 & located('Attacker Device', bluetoothZone, physical) \\ \hline
			11 & flowBind(statusUpdate, bluetooth, \_) \\ \hline
			14 & dataFlow('Generator3', 'PLC1', statusUpdate, oneWay) \\ \hline
			15-16 & crackPINCode(attacker, 'Generator3', 'PLC1') \\ \hline
			17-18 & \multirow{2}{0.9\textwidth}{crackPINCode(attacker, 'PLC1', 'Generator3')} \\
			17-39 & \\ \hline
			19 & isPairingProcess(pairingProcessPlc1Gen3) \\ \hline
			22 & flowBind(pairingProcessPlc1Gen3, bluetooth, \_) \\ \hline
			25 & dataFlow('PLC1', 'Generator3', pairingProcessPlc1Gen3, twoWay) \\ \hline
			26-27 & accessDataFlow(attacker, pairingProcessPlc1Gen3, view) \\ \hline
			28-29 & l2Connection('PLC1', 'Generator3', bluetoothZone, bluetooth, wireless) \\ \hline
			30 & existingProtocol(bluetoothZone, bluetooth) \\ \hline
			33 & located('Generator3', bluetoothZone, physical) \\ \hline
			36 & located('PLC1', bluetoothZone, physical) \\ \hline
			37 & inDiscoveryMode('Generator3') \\ \hline
			38 & malicious(attacker) \\ \hline
		\end{tabular}
	\end{subfigure}
	\caption{Bluetooth PIN cracking attack graph.}
    \label{fig:pin-crack-poc}
\end{figure}

In this scenario, the attacker is located in \textit{Bluetooth Zone} with \textit{Attacker Device} that is able to capture Bluetooth communication.
The attack graph generated for this scenario is presented in Figure \ref{fig:pin-crack-poc}. 
The attacker, who managed to place his/her device in \textit{Bluetooth Zone} (nodes 3-5 and 8), can view the communication between \textit{PLC\#1} and \textit{Generator\#3} (nodes 25-30, 33, 36-37). Given that this communication is the Bluetooth pairing process (nodes 19, 22), we can infer that the attacker can perform the PIN cracking attack (nodes 17-18).
After executing this attack, the attacker has the encryption key of the Bluetooth communication between \textit{PLC\#1} and \textit{Generator\#3} and can read their future communication (nodes 1-2, 11, 14).
Since the extracted key is symmetric, the encryption cracking also applies to the inverse communication direction (nodes 16-15, 39-17).

\subsection{\label{subsubsec:bus-dos-eval}Bus Denial of Service Attack}

To demonstrate attacks on the serial bus, we introduce the \textit{Serial Bus} component, which connects \textit{ PLC\#2}, two more generators (\textit{Generator\#4} and \textit{Generator\#5}), and a third PLC (\textit{PLC\#3}). The serial Modbus is the communication protocol in this bus, and the master is \textit{PLC\#2}.

\begin{figure}[t]
    \vspace{-30pt}
	\begin{subfigure}[h]{0.2\textwidth}
	    \includegraphics[scale=0.2]{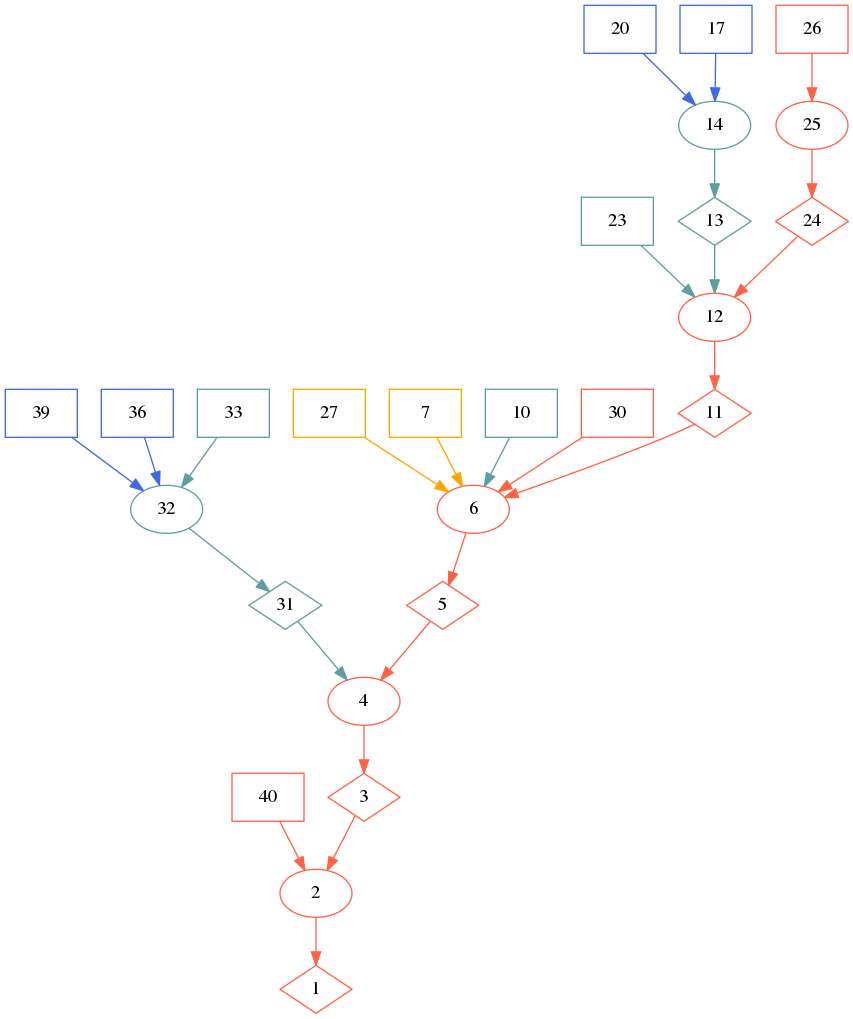}
	\end{subfigure}
	\hspace{50pt}
	\begin{subfigure}[h]{0.35\textwidth}
		\tiny
	    \vspace{120pt}
	    \hspace{-30pt}
		\begin{tabular}{|c|p{0.6\textwidth}|}\hline
			1-2 & dos(attacker, 'Generator4') \\ \hline
			3-4 & l2Access(attacker, 'PLC2', 'Generator4', modbus, 'Serial Bus', bus) \\ \hline
			5-6 & localAccess(attacker, 'PLC2', admin)  \\ \hline
			7 & isLoginService(ssh)  \\ \hline
			10 & aclH('PLC2', admin, 'HMI', 'PLC2',tcp, 22) \\ \hline
			11-12 & netAccess(attacker, 'HMI', 'PLC2', tcp, 22)  \\ \hline
			13-14 & aclNW('HMI', 'PLC2', tcp, 22)  \\ \hline
			17 & located('PLC2', 'OT Network', ipSubnet)  \\ \hline
			20 & located('HMI', 'OT Network', ipSubnet) \\ \hline
			23 & aclH('HMI', admin, 'HMI', 'PLC2', tcp, 22)  \\ \hline
			24-25 & localAccess(attacker, 'HMI', admin) \\ \hline
			26 & attackerLocated('HMI') \\ \hline
			27 & networkService('PLC2', ssh, tcp, 22, admin)  \\ \hline
			30 & hasAccount(attacker, 'PLC2', admin) \\ \hline
			31-32 & l2Connection('PLC2', 'Generator4', 'Serial Bus', modbus, bus)  \\ \hline
			33 & existingProtocol('Serial Bus', modbus)  \\ \hline
			36 & located('Generator4', 'Serial Bus', bus) \\ \hline
			39 & located('PLC2', 'Serial Bus', bus)  \\ \hline
			40 & malicious(attacker) \\ \hline
		\end{tabular}
	\end{subfigure}
	\hspace{-50pt}
	\caption{Bus denial of service attack graph.}
    \label{fig:bus-dos-poc}
\end{figure}

The attack graph generated for this attack scenario is presented in Figure \ref{fig:bus-dos-poc}. 
The attacker is located in \textit{HMI} (nodes 24-26), which can communicate with other components connected to \textit{OT Network}, in particular with \textit{PLC\#2} on port 22 (nodes 11-14, 17, 20, 23).
\textit{PLC\#2} enables remote login via SSH service (nodes 7, 27). The attacker that managed to obtain \textit{PLC\#2}'s admin account (node 30) can now connect to it on port 22, since inbound communication from \textit{HMI} is allowed (node 10), enabling the attacker to gain local access (nodes 5-6).

\textit{PLC\#2} and \textit{Generator\#4} can communicate via the bus using the Modbus protocol (nodes 31-33, 36, 39).
Since the attacker has local access to \textit{PLC\#2} (node 5), we infer that he/she can abuse it to access \textit{Generator\#4} via the bus (nodes 3-4) and thus deny service to \textit{Generator\#4} (nodes 1-2).

\subsection{Bus Spoofing Attacks}

\begin{figure*}[t]
    \hspace{-35pt}
    \begin{subfigure}[h]{0.5\textwidth}
       \begin{subfigure}[h]{0.2\textwidth}
    	    \includegraphics[scale=0.2]{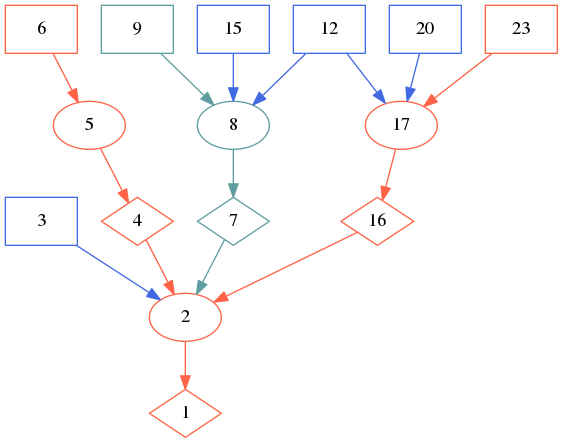}
    	\end{subfigure}
    	\hspace{75pt}
    	\begin{subfigure}[t]{\textwidth}
    		\tiny
    		\begin{tabular}{|c|p{0.45\textwidth}|}\hline
    			1-2 & spoofLinkHost(attacker, 'PLC2', 'Generator4', 'PLC3', deception) \\ \hline
    			3 & isMaster('PLC2', 'Serial Bus') \\ \hline
    			4-5 & localAccess(attacker, 'PLC3', admin) \\ \hline
    			6 & attackerLocated('PLC3') \\ \hline
    			7-8 & l2Connection('PLC3', 'Generator4', 'Serial Bus', modbus, bus) \\ \hline
    			9 & existingProtocol('Serial Bus', modbus) \\ \hline
    			12 & located('Generator4', 'Serial Bus', bus) \\ \hline
    			15 & located('PLC3', 'Serial Bus', bus) \\ \hline
    			16-17 & vulLinkProtocol('PLC2', 'Generator4', noAuthentication, modbus, adjacent, impersonateSrc) \\ \hline
    			20 & located('PLC2', 'Serial Bus', bus) \\ \hline
    			23 & vulLinkProtocol('Serial Bus', noAuthentication, modbus, adjacent, impersonateSrc) \\ \hline
    		\end{tabular}
    	\end{subfigure}
	\caption{Spoofing as bus master attack graph.}
    \label{fig:bus-spoof-master-poc}
    \end{subfigure}
    \hspace{30pt}
    \begin{subfigure}[t]{0.5\textwidth}
       \vspace{-66pt}
       \begin{subfigure}[h]{0.2\textwidth}
    	    \includegraphics[scale=0.2]{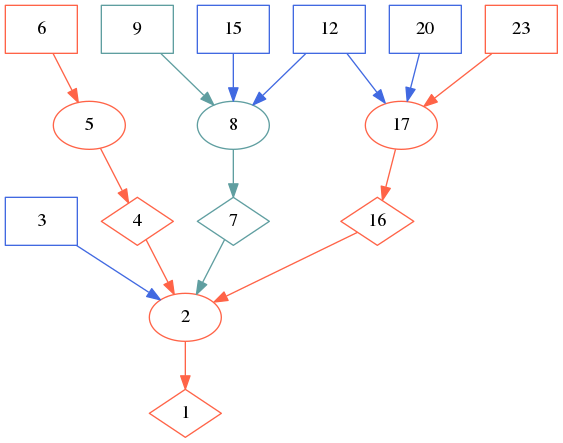}
    	\end{subfigure}
    	\hspace{75pt}
    	\begin{subfigure}[h]{\textwidth}
    	    \vspace{3pt}
    		\tiny
    		\begin{tabular}{|c|p{0.45\textwidth}|}\hline
    			1-2 & spoofLinkHost(attacker, 'Generator5', 'PLC2', 'PLC3', deception) \\ \hline
    			3 & isSlave('Generator5', 'Serial Bus') \\ \hline
    			4-5 & localAccess(attacker, 'PLC3', admin) \\ \hline
    			6 & attackerLocated('PLC3') \\ \hline
    			7-8 & l2Connection('PLC3', 'PLC2', 'Serial Bus', modbus, bus) \\ \hline
    			9 & existingProtocol('Serial Bus', modbus) \\ \hline
    			12 & located('PLC2', 'Serial Bus', bus) \\ \hline
    			15 & located('PLC3', 'Serial Bus', bus) \\ \hline
    			16-17 & vulLinkProtocol('Generator5', 'PLC2', noAuthentication, modbus, adjacent, impersonateSrc) \\ \hline
    			20 & located('Generator5', 'Serial Bus', bus) \\ \hline
    			23 & vulLinkProtocol('Serial Bus', noAuthentication, modbus, adjacent, impersonateSrc) \\ \hline
    		\end{tabular}
    	\end{subfigure}
	\caption{Spoofing as bus slave attack graph.}
    \label{fig:bus-spoof-slave-poc}
    \end{subfigure}
    \caption{Bus spoofing attack graphs.}
\end{figure*}

The Modbus protocol used in \textit{Serial Bus} is a master/slave communication protocol. In this type of communication, only one component (referred to as the master) is responsible for managing the operation of the other components (referred to as slaves) by sending command messages; the slaves only respond to the master's commands.

The first spoofing attack scenario is focused on impersonating the bus master. We demonstrate how an attacker located in \textit{PLC\#3} can exploit the lack of authentication in the serial Modbus protocol and impersonate the master of the bus (\textit{PLC\#2}) in order to send fake commands to \textit{Generator\#4}. The attack graph representing this scenario is presented in Figure \ref{fig:bus-spoof-master-poc}.

\textit{PLC\#3} and \textit{Generator\#4} can communicate via the bus (nodes 7-9, 12, 15). \textit{PLC\#2} and \textit{Generator\#4} can also communicate via the bus using Modbus, which lacks authentication (nodes 12, 16-17, 20, 23). The attacker can exploit his/her access to the bus via \textit{PLC\#3} (nodes 4-6, 15) and Modbus's vulnerability and impersonate \textit{PLC\#2} (the master) against \textit{Generator\#4} (nodes 1-3).

The second spoofing attack is focused on spoofing as a slave (\textit{Generator\#5}) in order to provide a fake response to the master (\textit{PLC\#2}). The attack graph (see Figure \ref{fig:bus-spoof-slave-poc}) is quite similar to the master spoofing scenario; the only difference is that we require the source of the vulnerable communication (node 16) to be a slave in the bus (node 3).

\section{Conclusion and Future Work}
In this paper, we introduce an extended network modeling for the MulVAL framework and evaluated it in an operational testbed simulating a simplified thermal power plant process.
The proposed modeling considers the physical network topology, supports short-range communication protocols, and models specific industrial communication architectures.
Furthermore, we model various network attacks, including bus spoofing, WEP cracking, Bluetooth PIN cracking, ARP spoofing, and DNS spoofing.

In future work, we plan to (1) extend the capabilities of the prototype system to support the automatic extraction of additional facts (such as obtaining information about wireless devices);
(2) model additional network protocols (such as Zigbee and BLE), wireless communication types (e.g., cellular communication) including the modeling of device mobility, and attack techniques; and (3) develop a process which utilizes the attack graph generated according to the extended modeling we presented in this paper to associate attack graph nodes with potential mitigation actions, resulting in a countermeasure plan that reduces the potential risk to the system.

\bibliographystyle{IEEEtran}
\bibliography{main}

\begin{thebibliography}{10}
\providecommand{\url}[1]{#1}
\csname url@samestyle\endcsname
\providecommand{\newblock}{\relax}
\providecommand{\bibinfo}[2]{#2}
\providecommand{\BIBentrySTDinterwordspacing}{\spaceskip=0pt\relax}
\providecommand{\BIBentryALTinterwordstretchfactor}{4}
\providecommand{\BIBentryALTinterwordspacing}{\spaceskip=\fontdimen2\font plus
\BIBentryALTinterwordstretchfactor\fontdimen3\font minus
  \fontdimen4\font\relax}
\providecommand{\BIBforeignlanguage}[2]{{%
\expandafter\ifx\csname l@#1\endcsname\relax
\typeout{** WARNING: IEEEtran.bst: No hyphenation pattern has been}%
\typeout{** loaded for the language `#1'. Using the pattern for}%
\typeout{** the default language instead.}%
\else
\language=\csname l@#1\endcsname
\fi
#2}}
\providecommand{\BIBdecl}{\relax}
\BIBdecl

\bibitem{phillips1998graph}
C.~Phillips and L.~P. Swiler, ``A graph-based system for network-vulnerability
  analysis,'' in \emph{Proceedings of the 1998 workshop on New security
  paradigms}.\hskip 1em plus 0.5em minus 0.4em\relax ACM, 1998, pp. 71--79.

\bibitem{sheyner2002automated}
O.~Sheyner, J.~Haines, S.~Jha, R.~Lippmann, and J.~M. Wing, ``Automated
  generation and analysis of attack graphs,'' in \emph{Security and privacy,
  2002. Proceedings. 2002 IEEE Symposium on}.\hskip 1em plus 0.5em minus
  0.4em\relax IEEE, 2002, pp. 273--284.

\bibitem{Jajodia2005topological}
S.~Jajodia, S.~Noel, and B.~O'Berry, \emph{Topological Analysis of Network
  Attack Vulnerability}.\hskip 1em plus 0.5em minus 0.4em\relax Springer US,
  2005, pp. 247--266.

\bibitem{ingols2006practical}
K.~Ingols, R.~Lippmann, and K.~Piwowarski, ``Practical attack graph generation
  for network defense,'' in \emph{Computer Security Applications Conference,
  2006. ACSAC'06. 22nd Annual}.\hskip 1em plus 0.5em minus 0.4em\relax IEEE,
  2006, pp. 121--130.

\bibitem{ou2006scalable}
X.~Ou, W.~F. Boyer, and M.~A. McQueen, ``A scalable approach to attack graph
  generation,'' in \emph{Proceedings of the 13th ACM conference on Computer and
  communications security}.\hskip 1em plus 0.5em minus 0.4em\relax ACM, 2006,
  pp. 336--345.

\bibitem{ou2005mulval}
X.~Ou, S.~Govindavajhala, and A.~W. Appel, ``Mulval: A logic-based network
  security analyzer.'' in \emph{USENIX Security Symposium}.\hskip 1em plus
  0.5em minus 0.4em\relax Baltimore, MD, 2005, pp. 8--8.

\bibitem{nvd}
``{NVD national vulnerability database},'' \url{http://www.nvd.nist.gov},
  [Online].

\bibitem{nessus}
``{Nessus security scanner},'' \url{http://www.nessus.org}, [Online].

\bibitem{acosta2016augmenting}
J.~C. Acosta, E.~Padilla, and J.~Homer, ``Augmenting attack graphs to represent
  data link and network layer vulnerabilities,'' in \emph{Military
  Communications Conference, MILCOM 2016-2016 IEEE}.\hskip 1em plus 0.5em minus
  0.4em\relax IEEE, 2016, pp. 1010--1015.

\bibitem{swiler2001computer}
L.~P. Swiler, C.~Phillips, D.~Ellis, and S.~Chakerian, ``Computer-attack graph
  generation tool,'' in \emph{discex}.\hskip 1em plus 0.5em minus 0.4em\relax
  IEEE, 2001, p. 1307.

\bibitem{ritchey2000using}
R.~W. Ritchey and P.~Ammann, ``Using model checking to analyze network
  vulnerabilities,'' in \emph{Security and Privacy, 2000. S\&P 2000.
  Proceedings. 2000 IEEE Symposium on}.\hskip 1em plus 0.5em minus 0.4em\relax
  IEEE, 2000, pp. 156--165.

\bibitem{ritchey2002representing}
R.~Ritchey, B.~O'Berry, and S.~Noel, ``Representing tcp/ip connectivity for
  topological analysis of network security,'' in \emph{Computer Security
  Applications Conference, 2002. Proceedings. 18th Annual}.\hskip 1em plus
  0.5em minus 0.4em\relax IEEE, 2002, pp. 25--31.

\bibitem{jha2002two}
S.~Jha, O.~Sheyner, and J.~Wing, ``Two formal analyses of attack graphs,'' in
  \emph{Proceedings 15th IEEE Computer Security Foundations Workshop. CSFW-15},
  June 2002, pp. 49--63.

\bibitem{ammann2002scalable}
\BIBentryALTinterwordspacing
P.~Ammann, D.~Wijesekera, and S.~Kaushik, ``Scalable, graph-based network
  vulnerability analysis,'' in \emph{Proceedings of the 9th ACM Conference on
  Computer and Communications Security}, ser. CCS '02.\hskip 1em plus 0.5em
  minus 0.4em\relax ACM, 2002, pp. 217--224. [Online]. Available:
  \url{http://doi.acm.org/10.1145/586110.586140}
\BIBentrySTDinterwordspacing

\bibitem{cve}
``{Common vulnerabilities and exposures dictionary.}''
  \url{http://www.cve.mitre.com}, [Online].

\bibitem{yi2013overview}
S.~Yi, Y.~Peng, Q.~Xiong, T.~Wang, Z.~Dai, H.~Gao, J.~Xu, J.~Wang, and L.~Xu,
  ``Overview on attack graph generation and visualization technology,'' in
  \emph{Anti-counterfeiting, security and identification (asid), 2013 IEEE
  international conference on}.\hskip 1em plus 0.5em minus 0.4em\relax IEEE,
  2013, pp. 1--6.

\bibitem{Jajodia2011cauldron}
S.~Jajodia, S.~Noel, P.~Kalapa, M.~Albanese, and J.~Williams, ``Cauldron
  mission-centric cyber situational awareness with defense in depth,'' in
  \emph{2011 - MILCOM 2011 Military Communications Conference}, Nov 2011, pp.
  1339--1344.

\bibitem{cauldron}
``{Cauldron, a cost-effective, nimble, adaptable and automated network
  visualization and modeling tool},'' \url{https://cyvision.net/cauldron/},
  [Online].

\bibitem{firemon}
``{Firemon},'' \url{http://firemon.com}, [Online].

\bibitem{skybox}
``{Skybox security},'' \url{http://www.skyboxsecurity.com}, [Online].

\bibitem{bacic2006mulval}
E.~Bacic, M.~Froh, and G.~Henderson, ``Mulval extensions for dynamic asset
  protection,'' CINNABAR NETWORKS INC OTTAWA (ONTARIO), Tech. Rep., 2006.

\bibitem{froh2009mulval}
M.~J. Froh and G.~Henderson, \emph{MulVAL extensions II}.\hskip 1em plus 0.5em
  minus 0.4em\relax Defence R \& D Canada-Ottawa, 2009.

\bibitem{ou2005logic}
X.~Ou and A.~W. Appel, \emph{A logic-programming approach to network security
  analysis}.\hskip 1em plus 0.5em minus 0.4em\relax Princeton University
  Princeton, 2005.

\bibitem{fluhrer2001weakness}
S.~Fluhrer, I.~Mantin, and A.~Shamir, ``Weaknesses in the key scheduling
  algorithm of rc4,'' in \emph{Selected Areas in Cryptography}, S.~Vaudenay and
  A.~M. Youssef, Eds.\hskip 1em plus 0.5em minus 0.4em\relax Berlin,
  Heidelberg: Springer Berlin Heidelberg, 2001, pp. 1--24.

\bibitem{vanhoef2017key}
M.~Vanhoef and F.~Piessens, ``Key reinstallation attacks: Forcing nonce reuse
  in wpa2,'' in \emph{Proceedings of the 2017 ACM SIGSAC Conference on Computer
  and Communications Security}.\hskip 1em plus 0.5em minus 0.4em\relax ACM,
  2017, pp. 1313--1328.

\bibitem{vanhoef2014advanced}
------, ``Advanced wi-fi attacks using commodity hardware,'' in
  \emph{Proceedings of the 30th Annual Computer Security Applications
  Conference}.\hskip 1em plus 0.5em minus 0.4em\relax ACM, 2014, pp. 256--265.

\bibitem{shaked2005cracking}
\BIBentryALTinterwordspacing
Y.~Shaked and A.~Wool, ``Cracking the bluetooth pin,'' in \emph{Proceedings of
  the 3rd International Conference on Mobile Systems, Applications, and
  Services}, ser. MobiSys '05.\hskip 1em plus 0.5em minus 0.4em\relax New York,
  NY, USA: ACM, 2005, pp. 39--50. [Online]. Available:
  \url{http://doi.acm.org/10.1145/1067170.1067176}
\BIBentrySTDinterwordspacing

\end{thebibliography}

\appendices

\onecolumn
\section{\label{app:prev-work-comp}Comparison Between MulVAL Extensions}
   \begin{table}[h!]
    \centering
    \scriptsize
    \begin{tabular}{|M{0.06\textwidth}|M{0.17\textwidth}|M{0.14\textwidth}|M{0.14\textwidth}|M{0.13\textwidth}|M{0.13\textwidth}|M{0.12\textwidth}|M{0.2\textwidth}|}
        \hline
         Work & Vulnerability Modeling & Host Modeling & Network Modeling & Data Modeling & User Modeling & Safeguard Modeling \\ 
         \hline
         Ou \etal \cite{ou2005mulval} (baseline) & 
          Characterized by exploitation range (\textit{local} or \textit{remote}) and consequence (impacting CIA, DoS, or privilege escalation); 
         Associated with a specific program running on a host
         &
         Characterized by running programs &
         Connectivity among hosts is represented by the \textit{hacl} predicate which is an abstraction of routing information and access controls &
         Associated with paths in a specific host and principals' access to files &
        User accounts to hosts and their privileges;
         Principal characteristic (malicious or incompetent)
         &
         -- \\
         \hhline{|=======|}
         Bacic \etal \cite{bacic2006mulval} & 
        -- &
        Model host classification (derived from the assets it contains) and software availability &
         Model subnets and inter-subnet communication &
         Model assets/data value (in terms of CIA) and classification &
         Model principal's clearance level &
         -- \\
         \hline
         Froh \etal \cite{froh2009mulval} & 
        -- &
        Model the value of IT services (in terms of CIA) and associate them with host and program &
         Associate hosts to networks;
         Represent network components (routers)
         &
         Model assets/data value (in terms of CIA)
         &
         -- &
        Represent security requirements (e.g., \textit{applicationAccount}) and incorporate them in the interaction rules; 
           Consider their (or other primitives) absence an indication to safeguard deployment
         \\
         \hline
         Liu \etal \cite{liu2015logic} & 
         Represent vulnerability exploitation based on evidence collected from hosts
        &
        -- &
         -- &
         -- &
         -- &
         -- \\
         \hline
         Acosta \etal \cite{acosta2016augmenting} & 
        Model ARP spoofing attack;
             Model null/weak authentication in OLSR
        &
        -- &
        Model network components (gateways;
         Model cross-subnet network access
         &
         Represent data flows (traffic generated by a user in a specific protocol that is exchanged between two hosts) &
         -- & 
         -- \\
         \hhline{|=======|}
         Our approach & 
          Represent protocol design vulnerabilities;
          Represent data vulnerabilities (e.g., lack of encryption);
          Model spoofing attacks, MITM, DoS, and attacks on wireless and bus communication
        &
        Represent local services &
         Represent hosts connectivity at different network layers; 
            Represent wireless and serial communications; 
           Distinguish between host and network-based firewall rules 
         &
         Represent specific data types (e.g., credentials); 
       Distinguish between data at rest and data in motion
         &
         Represent user access to hosts at different network layers
         &
         -- \\
         \hline
    \end{tabular}
    \label{tab:prev-work-comp}
\end{table}

\section{\label{app:mulval-example-input} Complete Input for the MulVAL Example}

\lstinputlisting[
  style      = Prolog-pygsty,
  float=hp,
  floatplacement=hp
]{mulval-example-input.pl}

\section{\label{app:fact-gen}Fact Generation Methods}
    \begin{table}[h!]
    \label{tab:facts-gen}
    \scriptsize
    \renewcommand*{\arraystretch}{1.3}
        \begin{tabular}{|c|c|m{0.66\textwidth}|}
            \hline
            \textbf{Fact} & \textbf{Generation method} & \begin{center}\textbf{Description}\end{center}
            \\ \hline
            isLoginService & Knowledge, Manual & Utilizing common knowledge or expert input to classify a service as a login service. For example: SSH and Remote Desktop are known services that enable users to log in to a machine.
            \\ \hline
            networkService & Assumption, Scan & Identifying network services directly from scan results. If the \textit{Protocol} or \textit{Port} arguments could not be detected, they are set to \textit{unknown}; if the \textit{User} argument is not detected, it is set to a wildcard.
            \\ \hline
            localService & Scan & Identify local services directly from host scans.
            \\ \hline
            isCredential & Assumption, Manual & Categorize data as containing credentials, for example: if some host runs a program that is defined as both \textit{isLoginService} and \textit{networkService}, an \textit{isCredential} fact is generated for each existing \textit{hasAccount} fact involving that host to indicate that the login data flow contains credentials (i.e., the remote login attempt): the \textit{User} and \textit{Host} arguments get the values of the \textit{User} and \textit{Host} arguments of \textit{hasAccount} predicate, and the \textit{DataName} argument is set to "\textit{user\_host}Credentials" (where \textit{user} and \textit{host} correspond to the \textit{User} and \textit{Host} arguments of \textit{hasAccount} predicate).
            \\ \hline
            dataFlow, flowBind & Knowledge, Manual & Generate data flows according to common knowledge, for example: the port number can provide an indication about the protocol used to send some data flow.
            \\ \hline
            dataBind & Assumption, Knowledge & Associating data with a specific path, for example: we know that a DNS server holds an IP to naming mapping in its cache, but we cannot get this memory mapping from a scan. 
            Thus, in order to represent this knowledge, this data is associated with a generic representation of this path.
            \\ \hline
            existingProtocol & Scan, Manual & Specify which protocol is used within different parts of the network (e.g., the usage of ARP within a subnet) directly from scan results or from expert's input.
            \\ \hline
            located & Scan & Identify the position of each device in the network directly from scan results.
            \\ \hline
            vulLinkProtocol, vulE2EProtocol & Knowledge & Associate protocol vulnerabilities to communications identified in the network by using a predefined knowledge base (e.g., NIST NVD).
            \\ \hline
            vulHost & Scan & Identify the vulnerabilities in each host directly from scan results (e.g., using Nessus).
            \\ \hline
            vulData & Manual & Associate vulnerabilities with data (e.g., unencrypted or unsinged) based on expert's input.
            \\ \hline
            hasAccount & Assumption, Scan & Associate user accounts to hosts directly from scans. If the \textit{Principal} argument could not be detected, it is set to: "\textit{host\_name}User" (where \textit{host\_name} is the value of the \textit{Host} argument); if the \textit{User} argument could not be detected, it is set to \textit{admin}.
            \\ \hline
            localAccess & Assumption, Scan & Infer a principal's access to hosts directly from scans. If the \textit{Principal} argument could not be detected, it is set as the \textit{hasAccount} predicate.
            \\ \hline
            aclNW & Assumption, Scan & Infer network-based firewall rules directly from scans. If a \textit{networkService} fact with an \textit{unknown} protocol and port was generated, a corresponding \textit{aclNW} with an \textit{unknown} protocol and port arguments is generated.
            \\ \hline
            aclH & Assumption, Scan & Infer host-based firewall rules directly from scans. Similar to \textit{aclNW}; if a network service with an unknown protocol and port was generated, a corresponding \textit{aclH} is generated, and the \textit{User} argument is set to a wildcard.
            \\ \hline
        \end{tabular}
\end{table}

\end{document}